\documentclass[12pt,a4paper]{article}

\usepackage{a4wide}
\usepackage{fancyhdr}
\usepackage{graphicx}
\usepackage{adjustbox}
\usepackage{diagbox}
\usepackage{caption}
\usepackage{subcaption}
\usepackage[ngerman, english]{babel}
\usepackage[utf8]{inputenc}
\usepackage[T1]{fontenc}
\usepackage{lmodern}
\usepackage{xcolor}
\usepackage[colorlinks,
citecolor={blue!70!black},
urlcolor={blue!70!black},
linkcolor={red!70!black},
hyperindex,breaklinks]{hyperref}
\usepackage{cite}
\usepackage{amsmath, amsthm, amssymb}
\usepackage{braket}
\usepackage{authblk}
\usepackage{tikz}
\usepackage{float}
\usepackage{indentfirst}
\usepackage[nodayofweek]{datetime}
\shortdate
\usepackage{orcidlink}

\begin{document}
%
%
%


\title{Thermalization in a model of enhanced memory capacity\thanks{~The present work is an improved version of chapter 4 ``Thermalization Despite Correlation'' of the author's Ph.D.\ thesis~\cite{Kaikov_2023}, and has neither been previously peer-reviewed nor published elsewhere.}
}


\author{Oleg Kaikov \orcidlink{0000-0002-9473-7294} \thanks{~\href{mailto:oleg.kaikov@cea.fr}{oleg.kaikov@cea.fr}} \\
{\small{\em Universit\'{e} Paris-Saclay, CEA, List, F-91120, Palaiseau, France}}\\
}

  


\date{\small{\today}}

\maketitle


\begin{abstract}
\noindent
We study thermalization within a quantum system with an enhanced capacity to store information. This system has been recently introduced to provide a prototype model of how a black hole processes and stores information. We perform a numerical finite-size analysis of this isolated quantum system and find indications that its information-carrying subsystem approaches thermality in the large system-size limit. The results lead us to suggest a novel thermalization mechanism. The corresponding distinguishing characteristic is that for a large class of physically meaningful non-equilibrium initial states $\ket{\text{in}}$, a few-body observable $\hat{A}$ thermalizes despite unignorable correlations between the fluctuations of its eigenstate expectation values $\bra{\alpha} \hat{A} \ket{\alpha}$ in the eigenstate basis of the model $\left\{ \ket{\alpha} \right\}$ and the fluctuations of the squared magnitudes of the coefficients $|C_{\alpha}|^2 = |\braket{\alpha | \text{in}} |^2$.
\end{abstract}


\section{Introduction}

\subsection{Overview}

Understanding how isolated quantum many-body systems thermalize is a longstanding area of research~\cite{Neumann_1929}. Specifically, consider such a system prepared in a pure initial state far from equilibrium. The question of interest is, what is the microscopic mechanism by which few-body observables within this system equilibrate under unitary time-evolution towards their typical expectation values given by an appropriate statistical ensemble? A distinguished mechanism for thermalization was established by the \emph{Eigenstate Thermalization Hypothesis} (ETH)~\cite{Deutsch_1991, Srednicki_1994}. This work has been further developed in~\cite{Srednicki_1994_2, Srednicki_1996, Srednicki_1999} among other studies (for reviews see e.g.~\cite{Nandkishore_2015, Gogolin_2016, DAlessio_2016, Mori_2018, Deutsch_2018}). Moreover, other mechanisms have been proposed, such as~\cite{Berges_2004, Rigol_2008, Biroli_2010, Ikeda_2011, Anza_2018, Li_2018}, among others.

The primary goal of this work is to test thermalization within a specific prototype system that exhibits an enhanced capacity to store information~\cite{Dvali_2018, Dvali_2020}. Furthermore, we would like to establish the microscopic mechanism that explains how this thermalization occurs. To investigate this, we consider finite-size realizations of the system. We find indications that certain few-body observables thermalize in the thermodynamic limit (TDL), i.e.\ in the limit of large system size. Moreover, our results suggest that this thermalization occurs via a new mechanism.

Specifically, a few-body observable $\hat{A}$ can still thermalize despite non-negligible correlations between two quantities: the fluctuations of its eigenstate expectation values $A_{\alpha \alpha} = \bra{\alpha} \hat{A} \ket{\alpha}$ in the eigenstate basis of the model $\{ \ket{\alpha} \}$, and the fluctuations of the squared magnitudes of the coefficients $|C_{\alpha}|^2 = |\braket{\alpha | \text{in}} |^2$ for a non-equilibrium initial state $\ket{\text{in}}$. The two sets of fluctuations are correlated in such a way that the infinite-time average of the few-body observable is equal to its statistical ensemble average.

The model~\cite{Dvali_2018, Dvali_2020} that we study was introduced to emulate the information-processing properties of a black hole. Other works, such as~\cite{Dvali_2013, Dvali_2015, Kaikov_2022}, have previously strengthened the connection of this and related systems to the original black hole fast scrambling conjecture~\cite{Hayden_2007, Sekino_2008}. However, our principal goal is more direct: We want to study thermalization in the specific realization~\cite{Dvali_2020} of a quantum system that exhibits an enhanced capacity to store information. Indeed, this quantum model is interesting in its own right and deserves an individual analysis.

In the remainder of the introduction we review the two aspects that are central for this work: thermalization and the enhanced memory capacity effect. In section~\ref{subsec_therm}, we briefly review thermalization in isolated quantum systems, the ETH, two other mechanisms that are relevant for the mechanism we introduce, as well as outline the new mechanism itself. In section~\ref{subsec_enh_mem_cap} we recapitulate the phenomenon of enhanced memory capacity as well as the model in~\cite{Dvali_2018, Dvali_2020}.

In section~\ref{sec_therm_test} we test the thermalization of certain few-body observables within the system by considering finite-size realizations of the model and extrapolating to the large-size limit. In section~\ref{sec_ETH_test} we test whether these few-body observables thermalize via the ETH. In sections~\ref{sec_test_other_mech} and~\ref{sec_TDC} we test other mechanisms in application to the thermalization of these few-body observables within the system. In section~\ref{sec_disc} we introduce the novel thermalization mechanism, discuss our findings and comment on possible applications. In section~\ref{sec_concl} we present our conclusions.

\subsection{Thermalization} \label{subsec_therm}

\subsubsection{Conditions for thermalization} \label{subsubsec_therm_cond}

Here we recapitulate thermalization of few-body observables within isolated quantum systems. Consider an isolated quantum many-body system of $D$ degrees of freedom. Let $\hat{H}$ denote the corresponding Hamiltonian. We denote its $\mathcal{N}$ energy eigenstates by $\ket{\alpha}$ and the corresponding energies by $E_{\alpha}$. Consider also a physically meaningful local few-body observable $\hat{A}$. Let $\hat{A}(t)=\mathrm{e}^{i \hat{H}t} \hat{A} \mathrm{e}^{-i \hat{H}t}$ be the corresponding Hermitian operator in the Heisenberg representation. Here and throughout the work we set $\hbar \equiv 1$. We prepare the system in an initial state $\ket{\text{in}}$, such that $\bra{\text{in}} \hat{A} \ket{\text{in}}$ is sufficiently far from the expectation value of $\hat{A}$ given by the microcanonical ensemble at the energy of the system $\bra{\text{in}} \hat{H} \ket{\text{in}}$. We say that the observable $\hat{A}(t)$ \emph{thermalizes} if~\cite{DAlessio_2016}:
\begin{itemize}
\item[(i)] The infinite-time average of the expectation value of the observable $\overline{\braket{\hat{A}(t)}}$ is equal to its microcanonical average, and
\item[(ii)] The temporal fluctuations of $\braket{\hat{A}(t)}$ about the microcanonical average are small at most later times.
\end{itemize}

\subsubsection{Conditions of the ETH} \label{subsubsec_ETH_cond}

We now review the ETH and state the corresponding conditions on the matrix elements of the observable in the eigenstate basis of the Hamiltonian. Specifically, the ETH states that an observable $\hat{A}(t)$ will thermalize if \cite{Deutsch_1991, Srednicki_1994, Srednicki_1996, Srednicki_1999, Rigol_2012, Anza_2018_2}:
\begin{itemize}
\item[(1)] The diagonal matrix elements $A_{\alpha \alpha} = \bra{\alpha} \hat{A} \ket{\alpha}$ vary approximately smoothly with $E_{\alpha}$ \emph{and} the magnitude of the difference between neighboring values $| A_{\alpha +1, \alpha +1} - A_{\alpha \alpha} |$ is exponentially small in $D$, and
\item[(2)] The magnitudes of the off-diagonal matrix elements $|A_{\alpha \beta}| = |\bra{\alpha} \hat{A} \ket{\beta}|$ with $\alpha \neq \beta$ are themselves exponentially small in $D$.
\end{itemize}

These are commonly expressed as~\cite{Srednicki_1996, Srednicki_1999, Murthy_2019, Murthy_2023, Berry_1977}
\begin{equation} \label{eqn_ETH}
A_{\alpha \beta} = \mathcal{A}(E)\delta_{\alpha \beta} + \mathrm{e}^{-S(E)/2}f_A(E,\omega)R_{\alpha \beta} \, .
\end{equation}
The average of the relevant energies is $E = (E_{\alpha} + E_{\beta})/2$ and their difference is $\omega = E_{\alpha} - E_{\beta}$. The microcanonical expectation value at $E$ is denoted by $\mathcal{A}(E)$. The thermodynamic entropy $S(E)$ at energy $E$ is given by the logarithm of the number of degenerate microstates. The function $f_A(E,\omega)$ is a real function corresponding to a particular observable $\hat{A}$, with $f_A(E,\omega)=f_A(E,-\omega)$. The functions $\mathcal{A}(E)$ and $f_A(E,\omega)$ vary smoothly with their arguments. The entries of the random Hermitian matrix $R_{\alpha \beta}$ have zero mean and unit variance.

As usual, the Hamiltonian is assumed to be non-degenerate~\cite{Srednicki_1996, Berry_1983}. Last, there is a condition that is commonly assumed to hold implicitly in the literature. Namely that the initial state of the system has a relatively small energy uncertainty~\cite{Srednicki_1994, Srednicki_1994_2, Srednicki_1996, Srednicki_1999, Rigol_2008, Rigol_2012}. We refer to this as the $0$-th condition of the ETH:

\begin{itemize}
\item[(0)] The initial state $\ket{\text{in}}$ is assumed to be a superposition of energy eigenstates, which are all sufficiently close in energy.
\end{itemize}

\subsubsection{Two related mechanisms} \label{subsubsec_Rigol_mech}

There are multiple examples in the literature of systems that achieve thermalization without satisfying the ETH conditions. This led to the development of thermalization mechanisms that are distinct from the ETH, such as~\cite{Berges_2004, Rigol_2008, Biroli_2010, Ikeda_2011, Anza_2018, Li_2018}. To provide for an informative link to the mechanism presented in this paper, we recapitulate two of the mechanisms considered in~\cite{Rigol_2008}. Consistently with the ETH condition (0), the spread in energy of the initial state is assumed to be sufficiently small. The two proposed mechanisms, that are different from the ETH and that lead to thermalization for initial states of physical interest, are~\cite{Rigol_2008}:
\begin{itemize}
\item[(i)] Both the eigenstate expectation values $A_{\alpha \alpha} = \bra{\alpha} \hat{A} \ket{\alpha}$ and the squared magnitudes of the coefficients $| C_{\alpha}|^2 = | \braket{\alpha | \text{in}} |^2$ exhibit large fluctuations, also for eigenstates that are close in energy. However, the fluctuations for these two quantities are not correlated. This characterizes the computation of the sum for the diagonal ensemble $\overline{\braket{\hat{A}(t)}} = \sum\limits_{\alpha=1}^{\mathcal{N}} |C_{\alpha}|^2 A_{\alpha \alpha}$. Namely, the sampling of the $A_{\alpha \alpha}$-s by the $| C_{\alpha}|^2$-s is unbiased.
\item[(ii)] There are essentially no fluctuations of the $| C_{\alpha}|^2$-s between eigenstates close in energy.
\end{itemize}
There is a further difference of these two mechanisms compared to the ETH~\cite{Rigol_2008}. Specifically, systems satisfying the ETH thermalize for all initial states that are narrow in energy. For the above two mechanisms, however, there may exist initial states that are narrow in energy but which do not lead to thermalization.

\subsubsection{A new mechanism}

There is a helpful connection that can be made between the mechanism introduced in this paper and the mechanism (i) of~\cite{Rigol_2008} (see section~\ref{subsubsec_Rigol_mech}). In both mechanisms the quantities $A_{\alpha \alpha}$ and $| C_{\alpha}|^2$ can exhibit large fluctuations. However, our mechanism allows for non-vanishing correlations between these sets of fluctuations, as long as the thermalization conditions (i) and (ii) from section~\ref{subsubsec_therm_cond} are fulfilled. The specific properties of the correlations are unimportant. The decisive characteristic is the following: We conduct a hypothesis test with the null hypothesis that the two sets of fluctuations are independent and an alternative hypothesis that they are not. If the test result is that the null hypothesis is to be rejected at a chosen significance level, this indicates that the two sets of fluctuations are correlated at that significance level.

In this work we perform a numerical analysis for finite-size realizations of the model in~\cite{Dvali_2018, Dvali_2020}. By extrapolation to the large system size limit, we find indications that certain few-body observables within this system thermalize in this limit (see section~\ref{sec_therm_test}). Furthermore, our results suggest that this occurs via a specific realization of the new mechanism outlined here (see section~\ref{subsec_new_therm_mech} for a detailed discussion).

We find indications that, for the considered few-body observables $\hat{A}$, the fluctuations in $A_{\alpha \alpha}$ and $| C_{\alpha}|^2$ persist in the TDL. Furthermore, our results suggest that these fluctuations remain correlated in this limit. For the finite-size realizations that we consider, we find that for the majority of the applied independence tests the independence hypothesis is rejected at the $0.05$ significance level. This leads us to propose that the correlations of the fluctuations are such that the corresponding biases collectively balance each other out and thermalization is achieved in agreement with the thermalization conditions of section~\ref{subsubsec_therm_cond}.

\subsection{Enhanced memory capacity} \label{subsec_enh_mem_cap}

The fundamental characteristic of the model in~\cite{Dvali_2018, Dvali_2020} is that of \emph{enhanced memory capacity}~\cite{Dvali_2018_2, Dvali_2018} (see e.g.~\cite{Dvali_2017, Dvali_2018_3, Dvali_2019, Dvali_2019_2, Dvali_2014, Dvali_2015, Dvali_2015_1, Dvali_2016, Alexandre_2024, Dvali_2024} for related work). Systems with this property allow for states with a high capacity to store information. The concept of enhanced memory capacity was introduced to model the information storing and processing properties of black holes, in alignment with the black hole's quantum $N$-portrait~\cite{Dvali_2013_2}.

An inherent effect of systems with an enhanced memory capacity is that of \emph{memory burden}~\cite{Dvali_2018, Dvali_2019}. This phenomenon takes place when the state of such a system stores a large amount of information. This information retains the system in a state of enhanced memory capacity and thereby stabilizes it. A system can achieve such a state by the means of \emph{assisted gaplessness}~\cite{Dvali_2019_2}. This occurs when a certain \emph{master mode} within the system becomes highly occupied. This master mode lowers the energy gaps of the other information-storing \emph{memory modes} through interaction. The memory modes can thereby store large amounts of information in terms of their occupation numbers. The memory modes backreact on the master mode via the memory burden effect and prevent it from rapidly losing its occupation number~\cite{Dvali_2018, Dvali_2019}. This stabilizes the system in a state of enhanced memory capacity. However, this can be avoided~\cite{Dvali_2018, Dvali_2020} if the system is able to rewrite the information stored in one set of memory modes onto a different set of memory modes. We review the corresponding prototype model below.

\subsubsection{The model} \label{subsubsec_model}

In (\ref{eqn_model}) we reproduce the prototype model with enhanced memory capacity, given in (34) of~\cite{Dvali_2020}, with minor changes in notation. We denote the model's two sets of bosonic memory modes by $K$ and $K'$. The creation and annihilation operators $\hat{a}_{k}^{\dagger}$, $\hat{a}_{k}$ and $\hat{a}_{k^{\prime}}^{\dagger}$, $\hat{a}_{k^{\prime}}$ of these modes satisfy the commutation relations
\begin{equation} \label{eqn_comm_rel}
\left[ \hat{a}_{j^{(\prime)}}, \hat{a}^{\dagger}_{k^{(\prime)}} \right] = \delta_{j^{(\prime)}k^{(\prime)}} \, , \quad \left[ \hat{a}_{j^{(\prime)}}, \hat{a}_{k^{(\prime)}} \right] = 0 \, , \quad \left[ \hat{a}^{\dagger}_{j^{(\prime)}}, \hat{a}^{\dagger}_{k^{(\prime)}} \right] = 0
\end{equation}
for $j^{(\prime)},k^{(\prime)}=1, \dotsc, K^{(\prime)}$. The corresponding occupation number operators are $\hat{n}_{k^{(\prime)}}=\hat{a}_{k^{(\prime)}}^{\dagger}\hat{a}_{k^{(\prime)}}$. The coupling $C_m$ sets the interaction strength between the memory modes $\hat{n}_{k^{(\prime)}}$. An additional bosonic mode $\hat{n}_a$, with creation and annihilation operators $\hat{a}^{\dagger}$ and $\hat{a}$, respectively, plays the role of the master mode. We denote a further bosonic mode, which exchanges occupation number with the mode $\hat{n}_a$, by $\hat{n}_b$, with creation and annihilation operators $\hat{b}^{\dagger}$ and $\hat{b}$, respectively. This interaction is controlled by the coupling $C_b$. The creation and annihilation operators $\hat{a}^{\dagger}$, $\hat{a}$ and $\hat{b}^{\dagger}$, $\hat{b}$ obey the commutation relations analogous to (\ref{eqn_comm_rel}).

The effective energy gaps of the memory modes $\hat{n}_k$ and $\hat{n}_{k'}$ are given by
\begin{equation}
\varepsilon_k \equiv \varepsilon \left( 1- \dfrac{n_a}{N} \right) \quad \text{and} \quad \varepsilon_{k'} \equiv \varepsilon \left( 1- \dfrac{n_a}{N-\Delta} \right) \, ,
\end{equation}
respectively. Their difference is controlled by the parameter $\Delta$. Let us remark that the total occupation number
\begin{equation}
N_m \equiv \sum\limits_{k=1}^K n_k + \sum\limits_{k'=1}^{K'} n_{k'}
\end{equation}
of the two memory sectors $K$ and $K'$ is conserved. According to~\cite{Dvali_2020}, we set the maximal occupation number of each memory mode to one. We denote the states of the system in the occupation number basis as
\begin{equation}
\ket{n_a, n_b, n_{k=1}, \dotsc, n_{k=K}, n_{k'=1}, \dotsc, n_{k'=K'}} \equiv \ket{n_a} \otimes \ket{n_b} \bigotimes_{k=1}^K \ket{n_k} \bigotimes_{k'=1}^{K'} \ket{n_{k'}} \, .
\end{equation}
We set the initial state of the system to
\begin{equation} \label{eqn_in_st}
\ket{\text{in}} = \ket{\underbrace{N}_{a}, \underbrace{0}_{b}, \underbrace{\overbrace{1, \dotsc, 1}^{=N_m}, 0, \dotsc, 0}_{K}, \underbrace{0, \dotsc, 0}_{K'}} \, .
\end{equation}
Here, the $\hat{n}_a$ and $\hat{n}_b$ modes contain $N$ and $0$ particles, respectively. Among the memory modes, the first $N_m$ modes of the $K$ memory sector are singly occupied, while all modes of the $K'$ sector are unoccupied. We follow~\cite{Dvali_2020} and fix the basic energy unit to $e \equiv 1$.

The Hamiltonian of the model is~\cite{Dvali_2020}
\begin{equation} \label{eqn_model}
\begin{aligned}
\hat{H} &= \varepsilon \left( 1 - \dfrac{\hat{n}_a}{N} \right) \sum\limits_{k=1}^{K} \hat{n}_k + \varepsilon \left( 1 - \dfrac{\hat{n}_a}{N-\Delta} \right) \sum\limits_{k'=1}^{K'} \hat{n}_{k'} + C_b \left( \hat{a}^{\dagger} \hat{b} + \text{H.c.} \right) \\
&+ C_m \left\lbrace \sum\limits_{k=1}^{K} \sum\limits_{k'=1}^{K'} f_1(k,k') \left( \hat{a}_k^{\dagger} \hat{a}_{k'} + \text{H.c.} \right) + \sum\limits_{k=1}^{K} \sum\limits_{l=k+1}^{K} f_2(k,l) \left( \hat{a}_k^{\dagger} \hat{a}_l + \text{H.c.} \right) \right. \\
&+ \left. \sum\limits_{k'=1}^{K'} \sum\limits_{l'=k'+1}^{K'} f_3(k',l') \left( \hat{a}_{k'}^{\dagger} \hat{a}_{l'} + \text{H.c.} \right) \right\rbrace \, .
\end{aligned}
\end{equation}
Here, the (essentially random) individual couplings of the memory modes are parametrized by
\begin{equation} \label{eqn_f_in_H}
f_i(k,l) = \begin{cases}
		   F_i(k,l)-1 \, , & F_i(k,l) < 0.5 \\
		   F_i(k,l) \, ,   & F_i(k,l) \geq 0.5
		   \end{cases}
\end{equation}
and
\begin{equation} \label{eqn_F_in_H}
F_i(k,l) = \left( \sqrt{2}(k+\Delta k_i)^3 + \sqrt{7}(l+\Delta l_i)^5 \right) \text{ mod } 1 \, ,
\end{equation}
where
\begin{equation} \label{eqn_Delta_in_H}
\Delta k_1 = \Delta k_2 = 1 \, , \quad \Delta k_3 = K+1 \, , \quad \Delta l_1 = \Delta l_3 = K+1 \, , \quad \Delta l_2 = 1 \, .
\end{equation}

In application to the prototype model in (\ref{eqn_model}), an argument of spherical symmetry for a static black hole given in~\cite{Dvali_2020} provides the constraint
\begin{equation}
\varepsilon = \sqrt{K}
\end{equation}
on the free energy gap $\varepsilon$ of the memory modes. Within this black hole prototype model, the scaling of the gravitational interaction with energy imposes a bound on the coupling $C_b$. In addition, retaining approximate gaplessness for the majority of the memory modes sets a bound on $C_m$. These bounds are given by~\cite{Dvali_2020}
\begin{equation} \label{eqn_CbCm_constr}
C_b \lesssim \dfrac{1}{\sqrt{N}} \quad \text{and} \quad C_m \lesssim \dfrac{1}{\sqrt{N_m}\sqrt{K}} \, ,
\end{equation}
respectively. Last, in the vicinity of the initial state (\ref{eqn_in_st}) the $K'$ memory sector should have a non-zero gap. This imposes the condition $|\varepsilon_{k'}| \gg 1/\sqrt{N_m}$~\cite{Dvali_2020}. This can equivalently be expressed as a condition on $\Delta$ as
\begin{equation} \label{eqn_DLT_constr}
\Delta \gg \dfrac{N}{1+\sqrt{N_m}\sqrt{K}} \, .
\end{equation}

As in~\cite{Dvali_2020}, we set $K=K'$. Furthermore, we select the values of the couplings at their respective limits (\ref{eqn_CbCm_constr}), namely $C_b = 1/\sqrt{N}$ and $C_m=1/(\sqrt{N_m} \sqrt{K})$. In addition, we scale the parameters of the system following the black hole's quantum $N$-portrait~\cite{Dvali_2013_2}. Specifically, for a black hole of entropy $S \gg 1$, we have $N=S$, $K=S$, $N_m=S/2$~\cite{Dvali_2020}. For the numerical simulations, if $N$ is odd, we round $N_m$ down to the nearest integer, $N_m=\lfloor N/2 \rfloor$. Last, we set $\Delta = N/2$. This fulfills the condition (\ref{eqn_DLT_constr}) for large $N$. The system is thus characterized by the only remaining free parameter, namely the system size $N$.

In sections~\ref{sec_therm_test}, \ref{sec_ETH_test}, \ref{sec_test_other_mech} and~\ref{sec_TDC} we perform numerical simulations with QuSpin~\cite{QuSpin_1, QuSpin_2, QuSpin_3} for $N \in [2, 9]$. We have numerically verified that there is no degeneracy in the eigenvalues $E_{\alpha}$ for each individual system that we consider.

As the ETH is applicable only within the considered symmetry sector of the system, we compute the total number of particles for the $\hat{n}_a$ and $\hat{n}_b$ modes, as well as for the memory modes, for all eigenstates $\ket{\alpha}$.\footnote{We thank Mari-Carmen Ba\~{n}uls for this comment.} We confirm, to within numerical accuracy, that all eigenstates have $N$ particles in total in the $\hat{n}_a$ and $\hat{n}_b$ sectors, and $N_m$ particles in total in the $K$ and $K'$ memory sectors, for all considered values of the system size $N$.

\begin{figure}[htbp]
    \centering
    \begin{subfigure}[b]{0.45\textwidth}
        \caption{\footnotesize $p(s)$ over $s$ for $N=8$}
        \includegraphics[width=\textwidth]{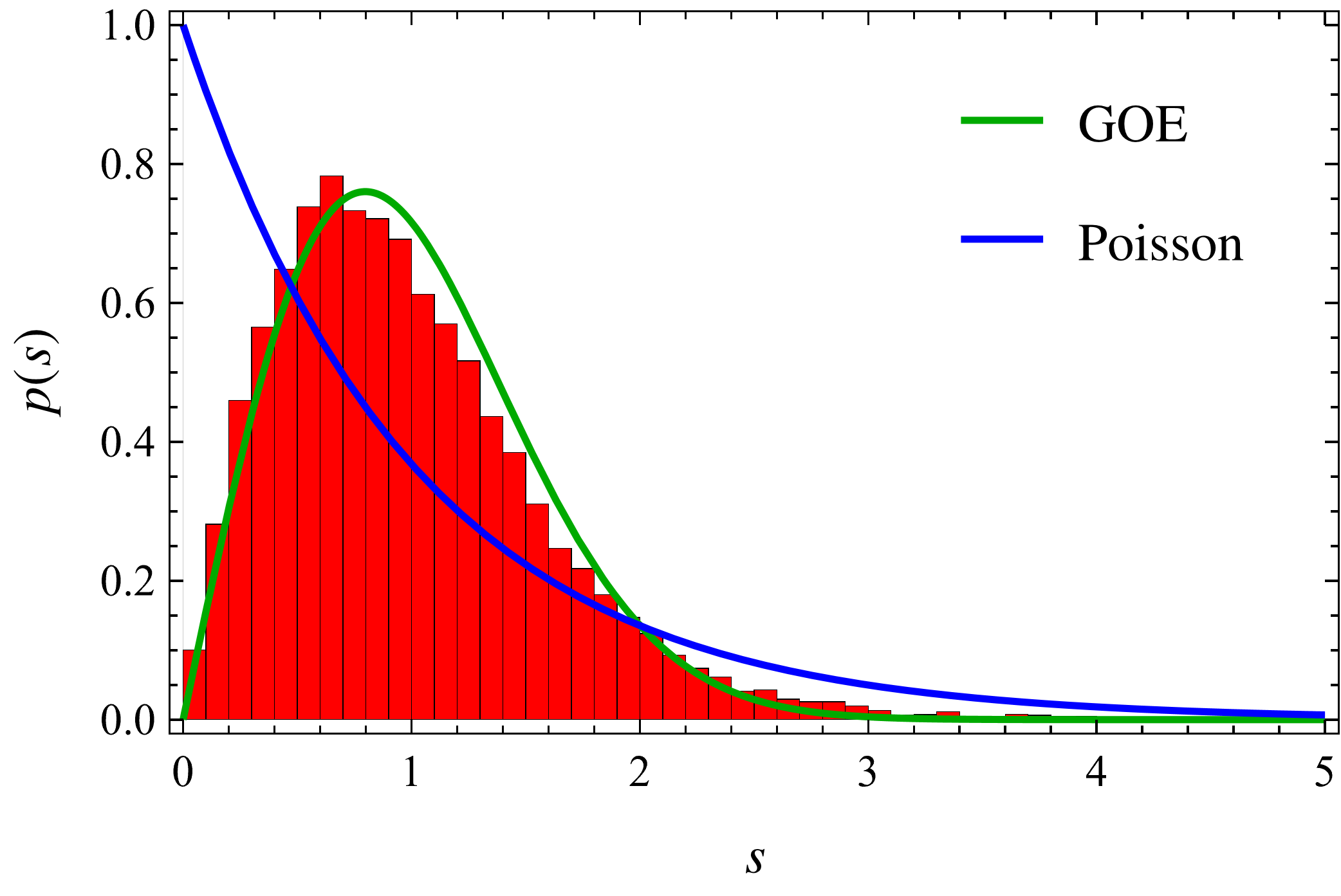}
        \centering
        \label{fig_lev_sp_dist_N_08}
    \end{subfigure}
    \begin{subfigure}[b]{0.45\textwidth}
    	\caption{\footnotesize $\operatorname*{arg\,max}\limits_{s} p(s)$ over $N$}
        \includegraphics[width=\textwidth]{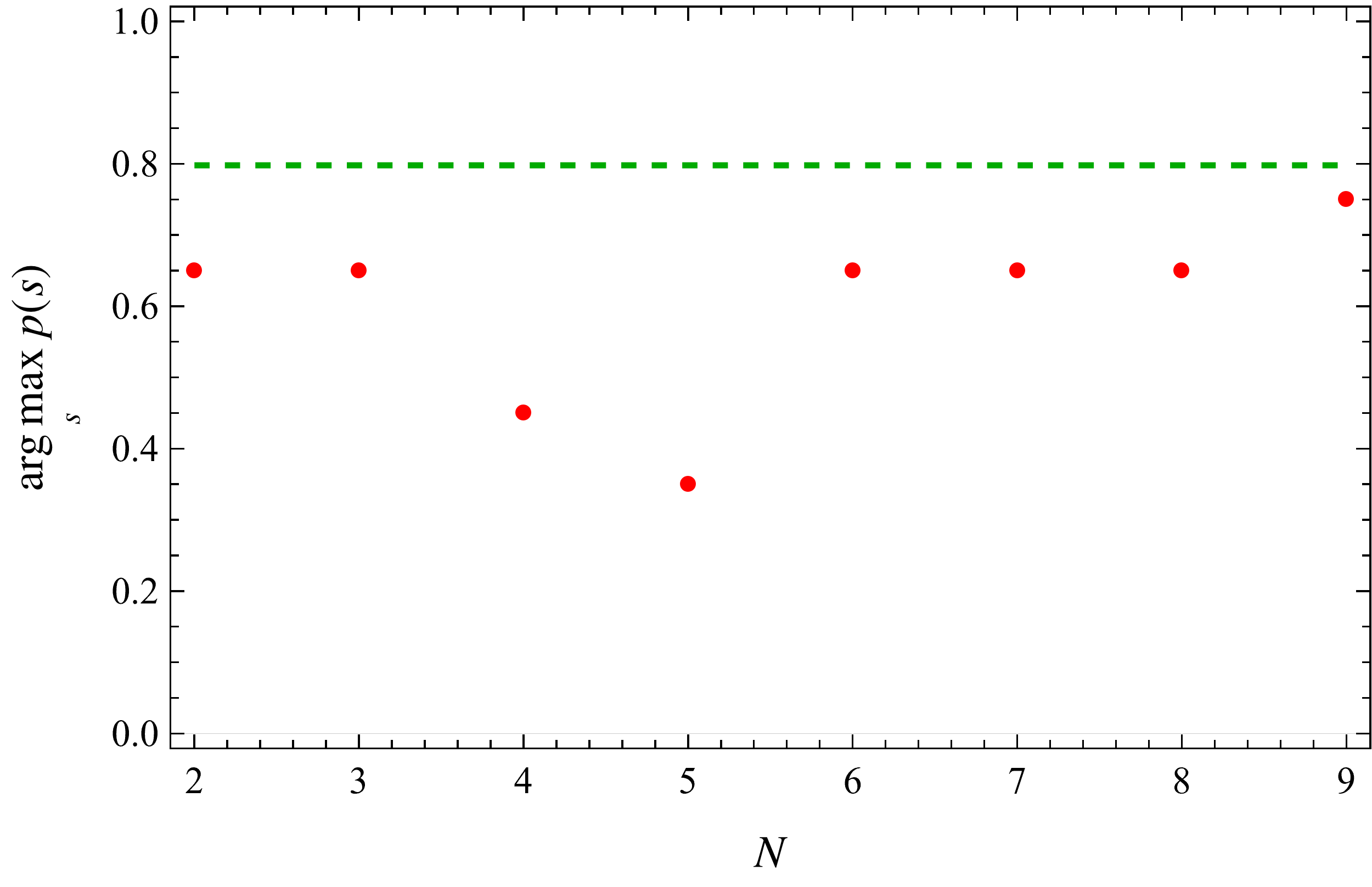}
        \label{fig_lev_sp_dist_argmax_vs_N}
    \end{subfigure}
    \caption{\small Level spacing distribution. Subfigure (\subref{fig_lev_sp_dist_N_08}) shows an example of the level spacing distribution $p(s)$ over the normalized level spacing $s$ for $N=8$. The histogram of the corresponding numerical data is shown in red. The solid lines represent the Wigner--Dyson (GOE) distribution (green) and the Poisson distribution (blue). Subfigure (\subref{fig_lev_sp_dist_argmax_vs_N}) displays the position of the maximum of $p(s)$ over $N$. The numerical data is shown as red points. The dashed green line represents the prediction of the GOE distribution, $\operatorname*{arg\,max}\limits_{s} p_{\text{GOE}}(s) = \sqrt{2/\pi} \approx 7.98 \times 10^{-1}$.}
    \label{fig_lev_sp}
\end{figure}

We also examine the level spacing statistics of the model.\footnote{We thank Maxim Olshanii for this suggestion and for sharing the corresponding code.} The results are shown in Fig.~\ref{fig_lev_sp}. For the considered level spacing distributions, all level spacings are normalized by the mean level spacing. Figure~\ref{fig_lev_sp_dist_N_08} shows the level spacing distribution $p(s)$ over the normalized level spacing $s$ for the model (\ref{eqn_model}) with $N=8$. We observe that the numerical level spacing distribution is well approximated by the Wigner--Dyson distribution, i.e.\ the Gaussian Orthogonal Ensemble (GOE)
\begin{equation}
p_{\text{GOE}}(s) = \dfrac{\pi}{2} s \, \mathrm{e}^{- \frac{\pi}{4} s^2}
\end{equation}
in this case. On the contrary, the Poisson distribution
\begin{equation}
p_{\text{Poisson}}(s) = \mathrm{e}^{-s}
\end{equation}
does not provide an adequate description of the data. Figure~\ref{fig_lev_sp_dist_argmax_vs_N} shows $\operatorname*{arg\,max}\limits_{s} p(s)$ over $N$. We observe that for increasing $N$, the position of the maximum of $p(s)$ of the numerical data approaches that of the GOE distribution, $\operatorname*{arg\,max}\limits_{s} p_{\text{GOE}}(s) = \sqrt{2/\pi} \approx 7.98 \times 10^{-1}$. This suggests that in the large-$N$ limit the level spacing statistics follows the Wigner--Dyson distribution. From the viewpoint of level spacing statistics, we therefore consider the model (\ref{eqn_model}) to be non-integrable in the TDL. For reviews on integrability see, for example,~\cite{Gogolin_2016, Caux_2011}.

\section{Test of thermalization} \label{sec_therm_test}

In this section we study whether the chosen few-body observables (see below) within the model (\ref{eqn_model}) satisfy the thermalization conditions of section~\ref{subsubsec_therm_cond}. For the model in (\ref{eqn_model}), the couplings of the individual memory modes are essentially random, and within each of the two memory sectors these can be relabeled. However, for each value of $N$, we examine a single instance of the system that is uniquely determined by (\ref{eqn_in_st}), (\ref{eqn_f_in_H}), (\ref{eqn_F_in_H}) and (\ref{eqn_Delta_in_H}). Considering multiple instances of the system and averaging over them risks smoothing out the non-thermalities. In the case when the considered sub-system does not thermalize, this could lead to falsely claiming that it does. Considering only one realization of the system avoids this undesired effect.

The couplings (\ref{eqn_f_in_H}) of the individual memory modes are essentially random on $[-1, 0.5) \cup [0.5, 1)$. We therefore assume that the chosen instance of the system is characteristic and captures the underlying physics. Based on the same arguments, we choose the occupation number $\hat{n}_i$ of a single memory mode $i=1$ from the $K$ sector as the few-body observable. Below we consider this observable and other quantities relevant to its thermalization for the system sizes $N \in [2, 9]$ and extrapolate our results to the large-$N$ limit.

\subsection{Studied quantities} \label{subsec_therm_test_quantities}

Here we define the quantities analyzed in the numerical simulations below. First, we reformulate the initial state (\ref{eqn_in_st}) in the eigenstate basis of the model $\{ \ket{\alpha} \}$ as
\begin{equation}
\ket{\text{in}} = \sum\limits_{\alpha=1}^{\mathcal{N}} C_{\alpha} \ket{\alpha} \, .
\end{equation}
Here, $\mathcal{N}$ is the dimension of the Hilbert space. The coefficients $C_{\alpha}$ are subject to the normalization condition $\sum\limits_{\alpha=1}^{\mathcal{N}} |C_{\alpha}|^2 = 1$. Having verified that the spectra of the individual model realizations do not have degeneracies, we can calculate the infinite-time average of the expectation value of the observable $\hat{n}_i$ to be
\begin{equation}
\overline{n}_i \equiv \overline{\braket{\hat{n}_i(t)}} = \lim\limits_{T \to \infty} \dfrac{1}{T} \int\limits_0^T \braket{\hat{n}_i(t)}\,\mathrm{d}t = \sum\limits_{\alpha=1}^{\mathcal{N}} |C_{\alpha}|^2 n_{i,\alpha \alpha} \, .
\end{equation}

During the time evolution, the time-dependent expectation value $\braket{\hat{n}_i(t)}$ exhibits temporal fluctuations about $\overline{n}_i$. The infinite-time root mean squared magnitude of these fluctuations can be expressed as
\begin{equation} \label{eqn_temp_fluct_sigma_i_t}
\sigma_{i,t} \equiv \left[ \overline{\braket{\hat{n}_i(t)}^2}-\overline{n}_i^2 \right]^{1/2} = \left[ \sum\limits_{\substack{\alpha, \beta \\ \alpha \neq \beta}} |C_{\alpha}|^2 |C_{\beta}|^2 |n_{i,\alpha\beta}|^2 \right]^{1/2} \, .
\end{equation}
The energy of the system, which we define as
\begin{equation}
\overline{E} \equiv \braket{\hat{H}} = \sum\limits_{\alpha=1}^{\mathcal{N}} |C_{\alpha}|^2 E_{\alpha} \, ,
\end{equation}
is conserved. Note from (\ref{eqn_in_st}) and (\ref{eqn_model}) that it is equal to zero, $\overline{E}=0$.

We now define several quantities to study the microcanonical ensemble average of $\hat{n}_i$. The quantum energy uncertainty of the system is given by
\begin{equation}
\sigma_{E,\text{q}} \equiv \left[ \braket{\hat{H}^2} - \braket{\hat{H}}^2 \right]^{1/2} = \left[ \sum\limits_{\alpha=1}^{\mathcal{N}} |C_{\alpha}|^2 \left( E_{\alpha} - \overline{E} \right)^2 \right]^{1/2} = \left[ \left( \sum\limits_{\alpha=1}^{\mathcal{N}} |C_{\alpha}|^2 E_{\alpha}^2 \right) - \overline{E}^2 \right]^{1/2} \, .
\end{equation}
We set the half-width of the energy window for the microcanonical ensemble average equal to $\sigma_{E,\text{q}}$~\cite{Srednicki_1994, Srednicki_1994_2, Srednicki_1996, Srednicki_1999, Rigol_2012}. We define $\mathcal{N}_{\sigma_{E,\text{q}}}$ as the number of eigenstates $\ket{\alpha}$ with eigenvalues $E_{\alpha}$ that lie within the interval $(\overline{E}-\sigma_{E,\text{q}},\overline{E}+\sigma_{E,\text{q}})$.

We cannot assess the smallness of the microcanonical energy window by comparing $\sigma_{E,\text{q}}$ to $\overline{E}$, since $\overline{E}=0$. However, note that the operator $(\hat{n}_a + \hat{n}_b)$ commutes with the Hamiltonian (\ref{eqn_model}). Adding this operator to the Hamiltonian, $\hat{H} \mapsto \hat{H} + (\hat{n}_a + \hat{n}_b)$, would therefore leave the dynamics of the system unaffected (see also the preliminary Hamiltonian in (16) of~\cite{Dvali_2020}). In this case, for the initial state (\ref{eqn_in_st}), the energy of the system would be equal to the system size, $\overline{E} \mapsto N$, while the quantum energy uncertainty of the system $\sigma_{E,\text{q}}$ would remain unchanged. Therefore, to judge the smallness of the energy window we instead compare $\sigma_{E,\text{q}}$ to the system size $N$ by considering $\sigma_{E,\text{q}}/N$.

Last, the microcanonical ensemble average of $\hat{n}_i$ is given by
\begin{equation}
n_{i,\text{mc}} \equiv \dfrac{1}{\mathcal{N}_{\sigma_{E,\text{q}}}} \sum\limits_{\substack{\alpha \\ |\overline{E} - E_{\alpha}| < \sigma_{E,\text{q}}}} n_{i,\alpha \alpha} \, .
\end{equation}

\subsection{Numerical analysis} \label{subsec_therm_test_num_an}

In this section we present the numerical results concerning the test of thermalization for the observable $\hat{n}_1$. For sufficiently small values of $\Delta$, we expect the total occupation number $N_m$ of the memory modes to become evenly distributed over the memory modes of both sectors $K$ and $K'$ for later times. That is, we expect $\braket{\hat{n}_1(t)}$ to equilibrate towards $\overline{n}_1 = \lfloor N/2 \rfloor / (2N)$ for $N_m=\lfloor K/2 \rfloor$ and $K=K'=N$, which simplifies to $\overline{n}_1 = 0.25$ for even values of $N$ as well as in the limit $N \to \infty$. We study this below.

\begin{figure}[htbp]
    \centering
    \includegraphics[width=0.5\textwidth]{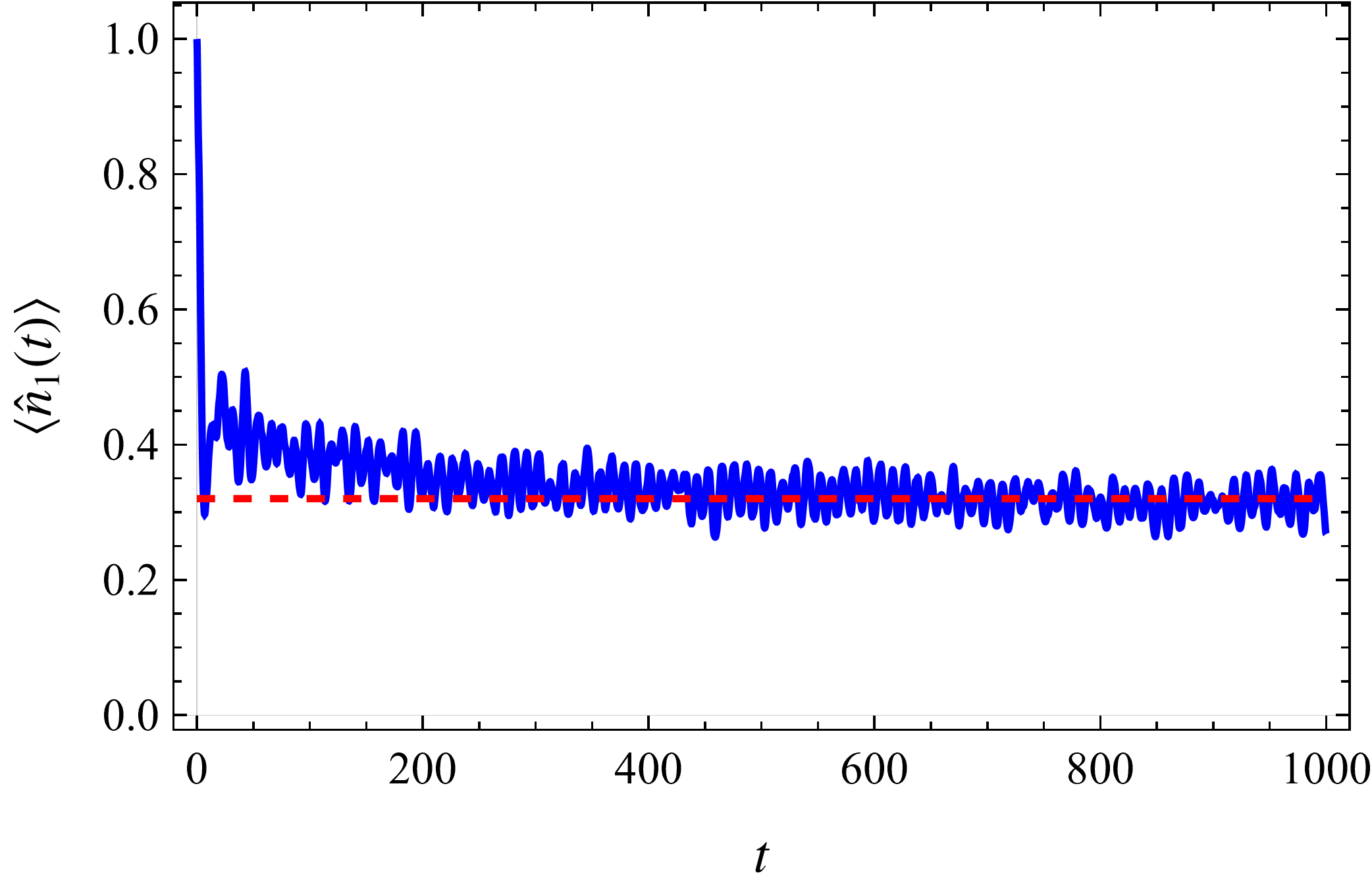}
    \caption{\small An example of relaxation dynamics for $N=8$. A plot of $\braket{\hat{n}_1(t)}$ over $t$ is shown in solid blue. The corresponding infinite-time average $\overline{n}_1 \approx 0.32$ is shown in dashed red.}
    \label{fig_ex}
\end{figure}

We first consider the relaxation dynamics of $\hat{n}_1$ for a realization of the system for $N=8$. The plot of $\braket{\hat{n}_1(t)}$ over time $t$ is shown in Fig.~\ref{fig_ex}. We observe that $\braket{\hat{n}_1(t)}$ equilibrates to $\overline{n}_1 \approx 0.32$. The temporal fluctuations of $\braket{\hat{n}_1(t)}$ about $\overline{n}_1$ are small at later times. Clearly, $\braket{\hat{n}_1(t)}$ does not thermalize, since this would require $\overline{n}_1 = n_{1,\text{mc}}=0.25$. We find analogous behavior of $\braket{\hat{n}_1(t)}$ for other values of $N$.

The regime $N \gg 1$ is the intended domain of applicability of the model (\ref{eqn_model}). Therefore, we are interested in the thermalization properties of the system in the TDL. We study this numerically, by extrapolating the results for finite-$N$ realizations of the system to the large-$N$ limit. We expect $\overline{n}_1$ and $n_{1,\text{mc}}$ to have the same asymptotic dependence on $N$ and to be equal to $0.25$ in the large-$N$ limit.

\begin{figure}[htbp]
    \centering
    \begin{subfigure}[b]{0.32\textwidth}
        \caption{\footnotesize $\overline{n}_1 \sim \mathrm{e}^{(-0.14 \pm 0.32) N}$}
        \includegraphics[width=\textwidth]{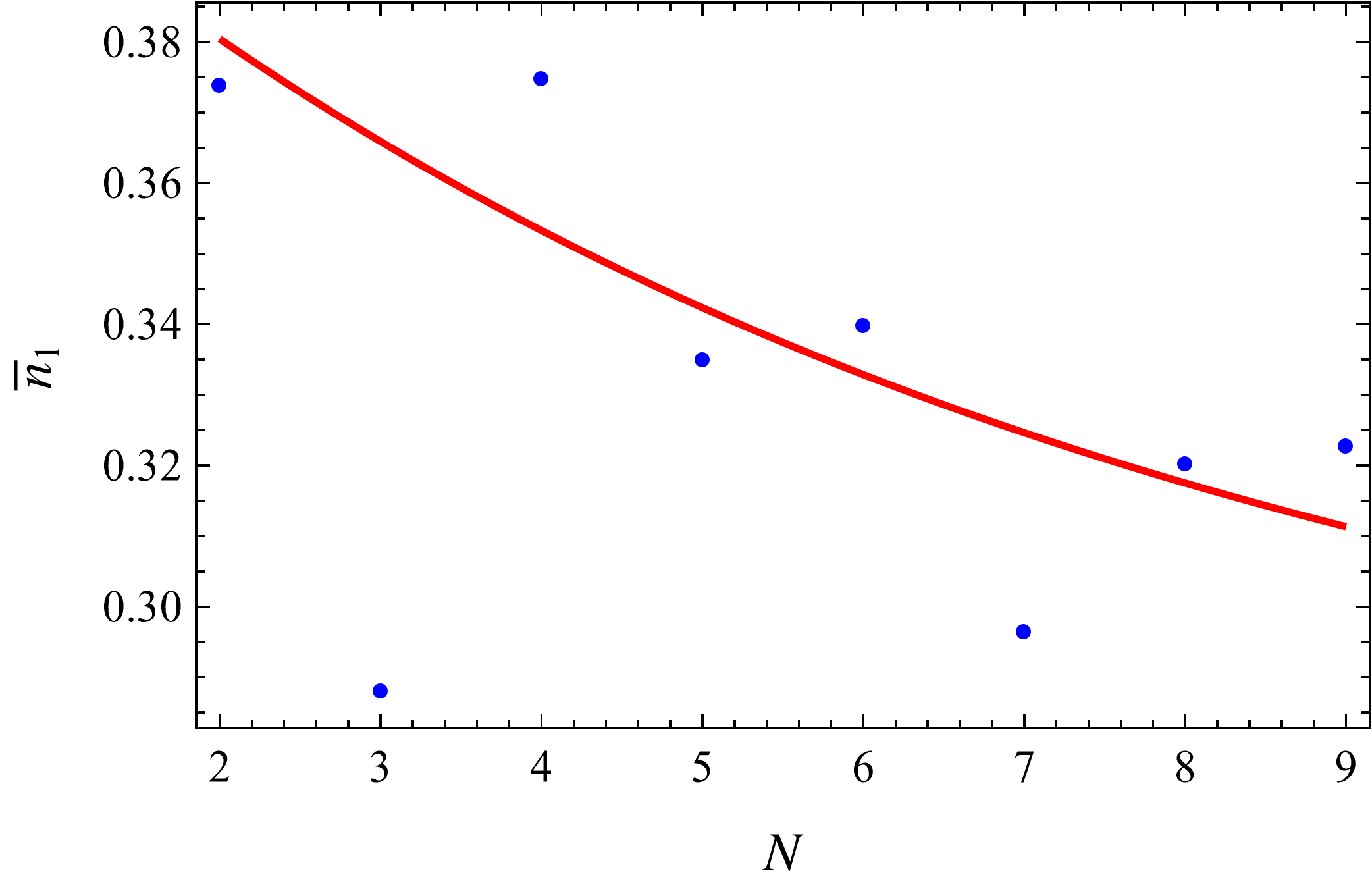}
        \centering
        \label{fig_nbar0_vs_N}
    \end{subfigure}
    \begin{subfigure}[b]{0.32\textwidth}
    	\caption{\footnotesize $n_{1,\text{mc}} \sim \mathrm{e}^{(-0.25 \pm 0.16) N}$}
        \includegraphics[width=\textwidth]{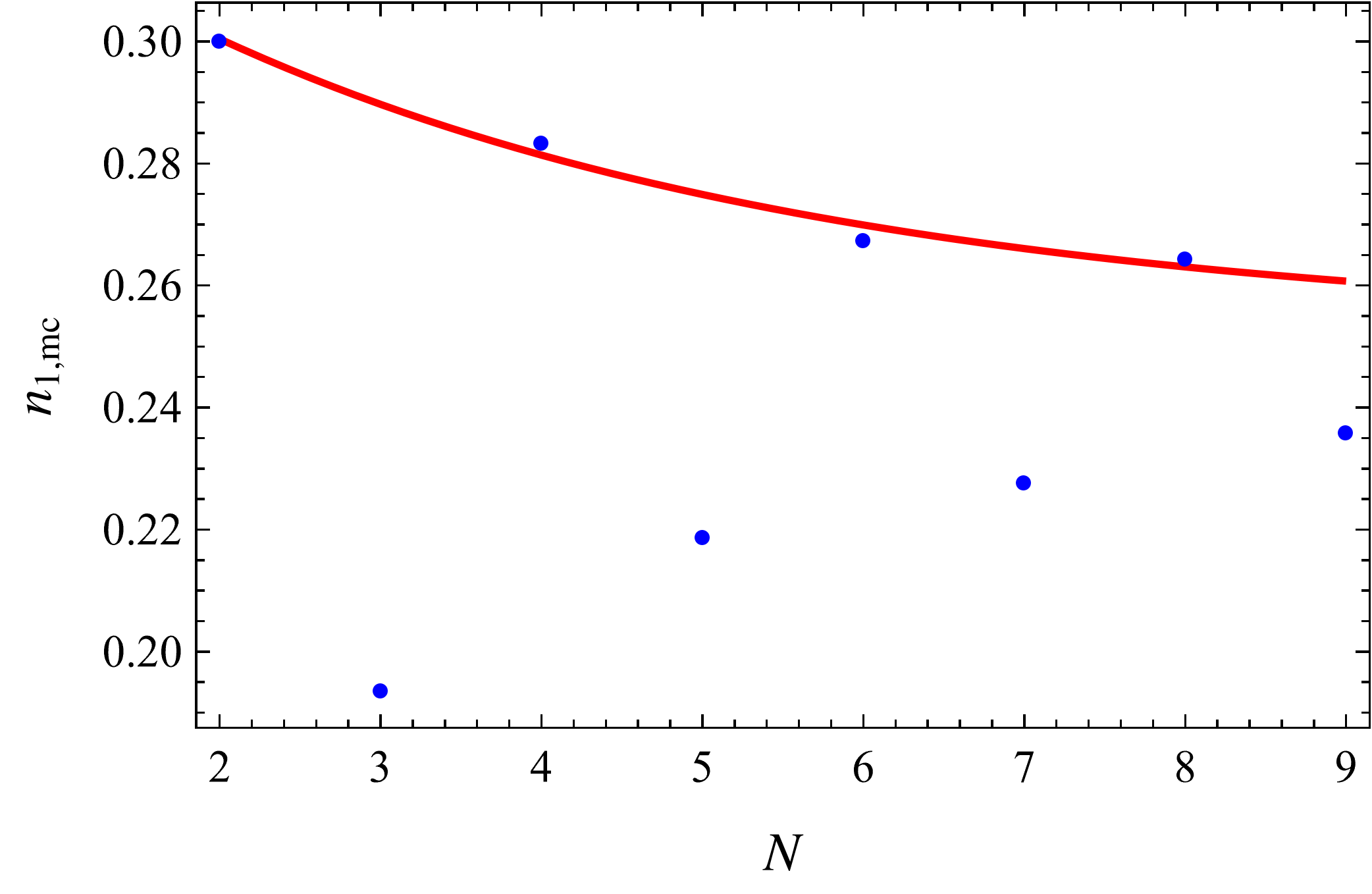}
        \label{fig_n0mc_vs_N}
    \end{subfigure}
    \begin{subfigure}[b]{0.32\textwidth}
    	\caption{\footnotesize $\sigma_{1,t} \sim \mathrm{e}^{(-0.470 \pm 0.074)N}$}
        \includegraphics[width=\textwidth]{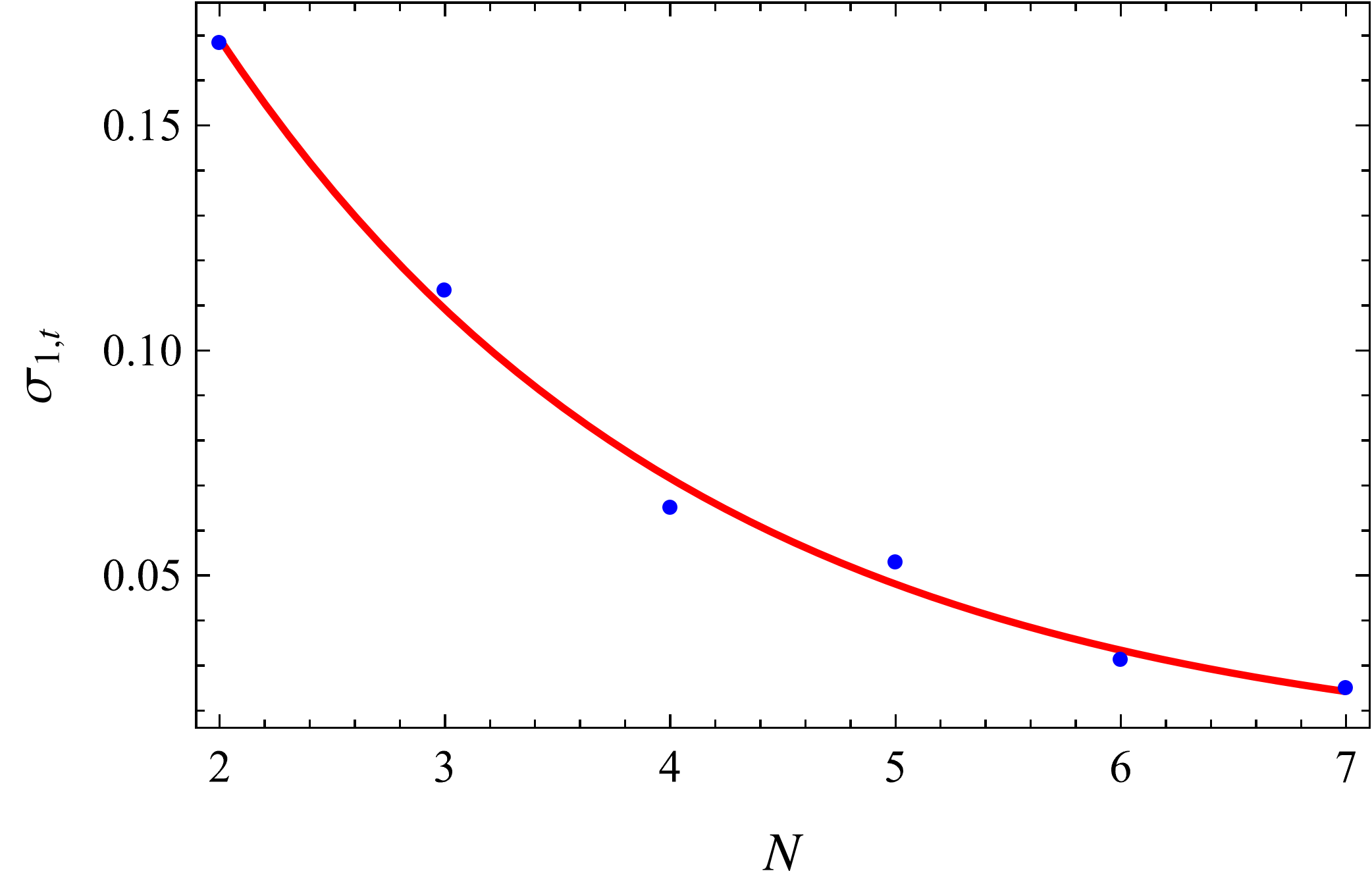}
        \label{fig_sig0t_vs_N}
    \end{subfigure}
    \begin{subfigure}[b]{0.32\textwidth}
    	\caption{\footnotesize $\frac{\sigma_{E,\text{q}}}{N} \sim N^{-0.900 \pm 0.033}$}
        \includegraphics[width=\textwidth]{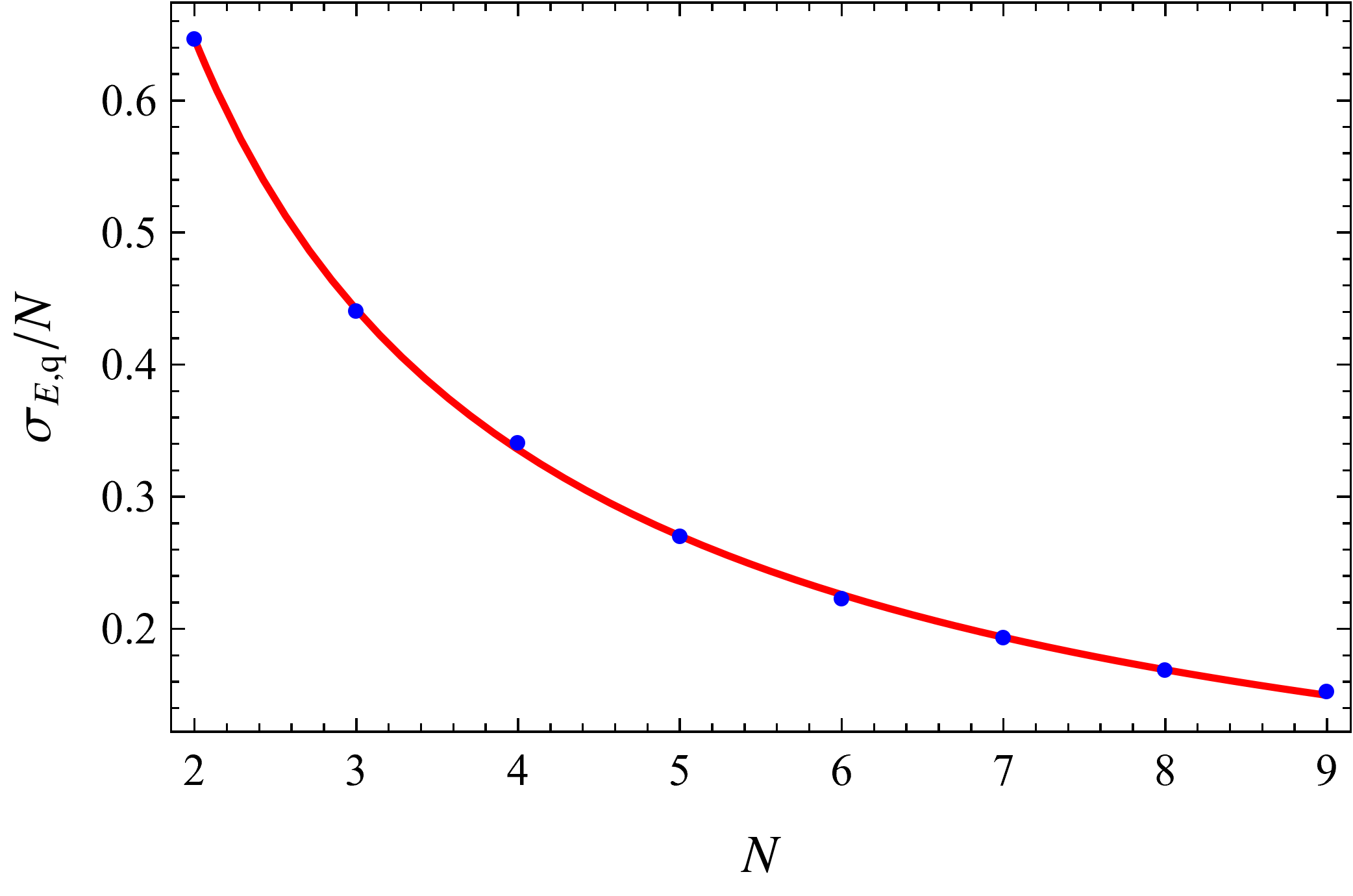}
        \label{fig_sigEq_o_N_vs_N}
    \end{subfigure}
    \begin{subfigure}[b]{0.32\textwidth}
    	\caption{\footnotesize $\mathcal{N}_{\sigma_{E,\text{q}}} \sim \mathrm{e}^{(0.71 \pm 0.12)N}$}
        \includegraphics[width=\textwidth]{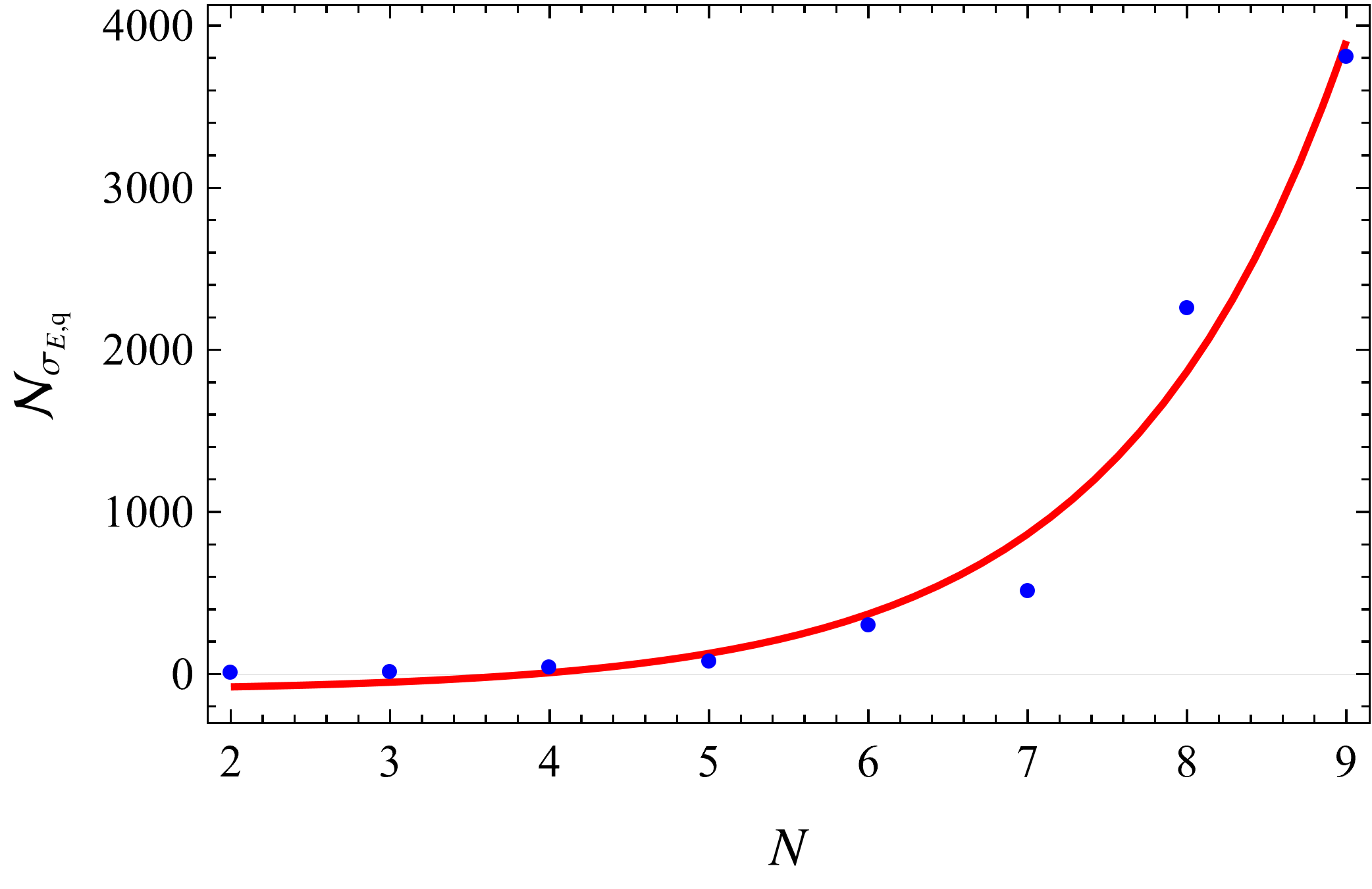}
        \label{fig_NsigEq_vs_N}
    \end{subfigure}
    \caption{\small Indication of thermalization in the large-$N$ limit. The quantities in section~\ref{subsec_therm_test_quantities} are plotted over $N$ in the individual subfigures. For compactness, the subcaptions show only the $N$-scaling of the corresponding quantities. The fit parameter values are presented in Table~\ref{tab_fit} in the Appendix. The numerical data is shown with points. The respective fits are represented as solid lines. The subfigures display: (\subref{fig_nbar0_vs_N}) the infinite-time average $\overline{n}_1$ of $\hat{n}_1$, (\subref{fig_n0mc_vs_N}) the microcanonical average $n_{1,\text{mc}}$ of $\hat{n}_1$, (\subref{fig_sig0t_vs_N}) the measure $\sigma_{1,t}$ of temporal fluctuations of $\braket{\hat{n}_1(t)}$ about $\overline{n}_1$, (\subref{fig_sigEq_o_N_vs_N}) the ratio of the quantum energy uncertainty to the system size $\sigma_{E,\text{q}}/N$, (\subref{fig_NsigEq_vs_N}) the number of eigenvalues $\mathcal{N}_{\sigma_{E,\text{q}}}$ within the interval $(\overline{E}-\sigma_{E,\text{q}},\overline{E}+\sigma_{E,\text{q}})$. The data point for $N=3$ was excluded from the fits of $\overline{n}_1(N)$. Only the data points with even $N$ were considered in the fits of $n_{1,\text{mc}}(N)$.}
    \label{fig_therm}
\end{figure}

Figure~\ref{fig_therm} shows the results for the quantities in section~\ref{subsec_therm_test_quantities} with the corresponding best obtained fit functions. The values of the respective fit parameters are given in Table~\ref{tab_fit} in the Appendix. In Table~\ref{tab_fit}, if multiple fits were performed for a quantity, the fits shown in the figures throughout the paper, along with the respective values of the fit parameters, are marked with a ``$^*$''. For completeness and to facilitate the comparison, we provide additional fits in Table~\ref{tab_fit} that may also be suitable for the obtained data. These fit functions may become more appropriate if the data for larger values of $N$ is obtained.

The small-$N$ outlier data point for $N=3$ was excluded from the fits for $\overline{n}_1$. We exclude the points with odd $N$ from the fit of $n_{1,\text{mc}}$. For these points, $N_m$ is given by $(N-1)/2$ instead of $N/2$. For small values of $N$ these points can misrepresent the $N$-dependence of the system.

The fit results of $\overline{n}_1$ and $n_{1,\text{mc}}$ in Table~\ref{tab_fit} (the corresponding quantities are shown in Figs.~\ref{fig_nbar0_vs_N} and~\ref{fig_n0mc_vs_N}, respectively) indicate that the thermalization condition (i) in section~\ref{subsec_therm}, i.e.\ $\overline{n}_1 = n_{1,\text{mc}} = 0.25$, can be fulfilled in the large-$N$ limit, if $\overline{n}_1$ and $n_{1,\text{mc}}$ scale as $\exp(-cN)$ with $N$, where $c$ is a positive constant. Figure~\ref{fig_n0mc_vs_N} suggests that $n_{1,\text{mc}}$ for both even and odd values of $N$ converges to $0.25$. We therefore find evidence that the thermalization condition (i) in section~\ref{subsec_therm} is satisfied in the large-$N$ limit. For finite $N$, the deviations from ideal thermality are exponentially small in $N$.

The infinite-time root mean squared magnitude of the temporal fluctuations $\sigma_{1,t}$ of $\braket{\hat{n}_1(t)}$ about $\overline{n}_1$ is shown in Fig.~\ref{fig_sig0t_vs_N}. The corresponding fit result in Table~\ref{tab_fit} provides evidence that $\sigma_{1,t}$ approaches zero exponentially with $N$ in the large-$N$ limit. This indicates that the thermalization condition (ii) in section~\ref{subsec_therm} is satisfied. Based on the numerical results for the thermalization conditions (i) and (ii) from section~\ref{subsec_therm}, we therefore conclude that the observable $\hat{n}_1$ thermalizes in the TDL.

The fit result in Table~\ref{tab_fit} suggests that $\sigma_{E,\text{q}}/N$ vanishes in the large-$N$ limit (see Fig.~\ref{fig_sigEq_o_N_vs_N} for the corresponding plot). That is, we find indications that the quantum energy uncertainty of the system is much smaller than its energy (for the ``shifted'' Hamiltonian $\hat{H} \mapsto \hat{H} + (\hat{n}_a + \hat{n}_b)$) for large $N$. Moreover, we obtain that $\sigma_{E,\text{q}}/N \sim N^{-0.900 \pm 0.033}$, i.e.\ that $\sigma_{E,\text{q}}$ is algebraically small in $N$ when compared to the ``shifted'' system energy $\overline{E} \mapsto N$. This is typically assumed to hold for states of physical interest, see e.g.\ $\sigma_{E,\text{q}} \sim N^{-1/2} \overline{E}$ in~\cite{Srednicki_1996, Srednicki_1999, Rigol_2012}. Thus, the fit result of $\sigma_{E,\text{q}}/N$ suggests that the microcanonical energy window becomes infinitesimally small in the TDL.

Last, we remark that the number of eigenvalues $\mathcal{N}_{\sigma_{E,\text{q}}}$ within the interval $(\overline{E}-\sigma_{E,\text{q}},\overline{E}+\sigma_{E,\text{q}})$ increases with $N$, as can be seen from Fig.~\ref{fig_NsigEq_vs_N} and the corresponding fit results in Table~\ref{tab_fit}.

\section{Test of ETH} \label{sec_ETH_test}

To test the ETH for the observable $\hat{n}_1$ within the model (\ref{eqn_model}), we need to check the ETH conditions (0), (1) and (2) of section~\ref{subsubsec_ETH_cond}. The following numerical analysis indicates that $\hat{n}_1$ does not satisfy the ETH condition (1) on the diagonal matrix elements $n_{1,\alpha\alpha}$. We elaborate on this and the other conditions below.

\subsection{Condition (0)}

To satisfy the ETH condition (0), the initial state should have a sufficiently small energy uncertainty. In other words, the initial state must be a superposition of eigenstates that are all sufficiently close in energy.

\begin{figure}[htbp]
    \centering
    \includegraphics[width=0.5\textwidth]{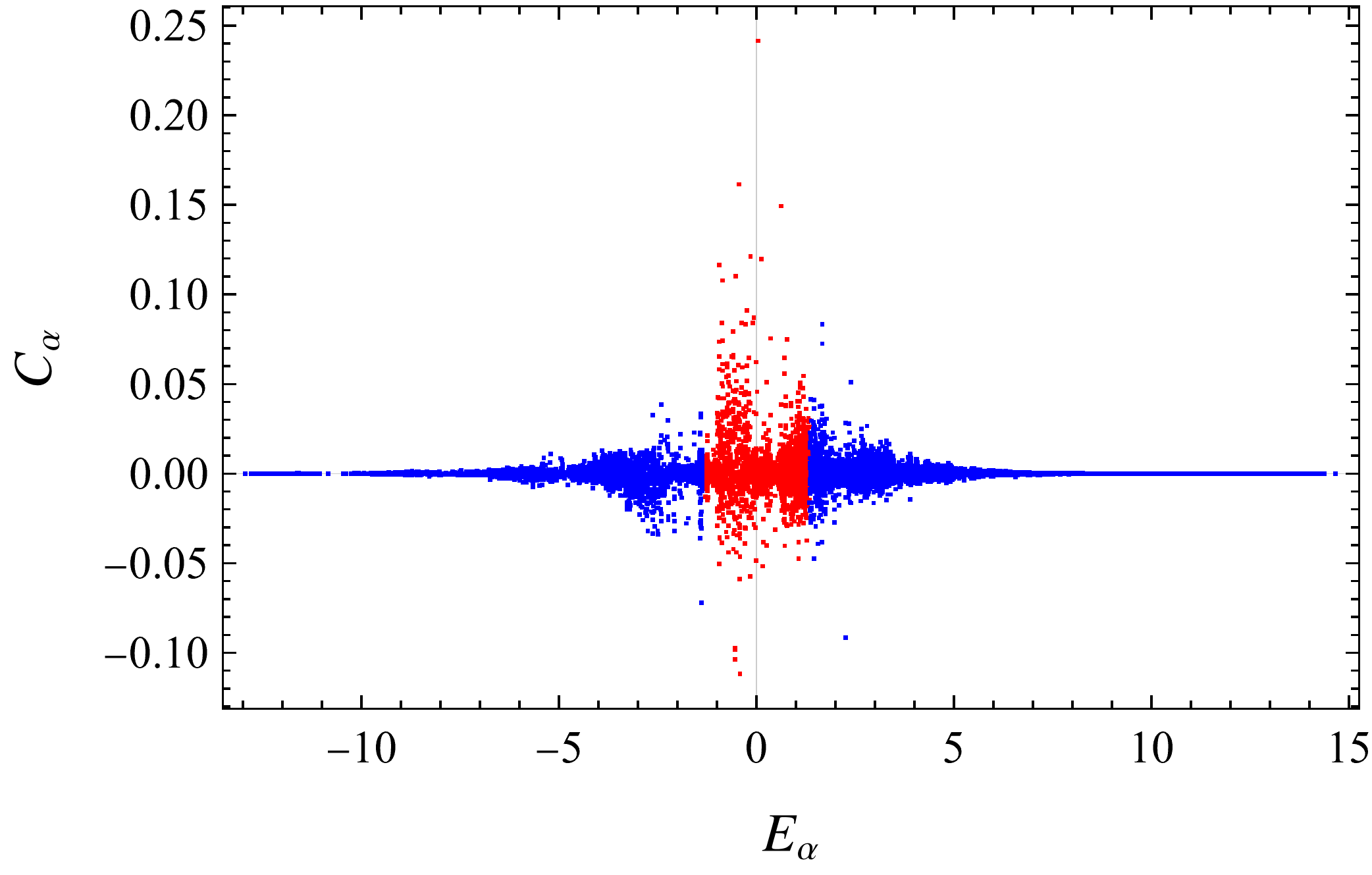}
    \caption{\small A plot of the coefficients $C_{\alpha}$ over the corresponding eigenstate energies $E_{\alpha}$ for $N=8$. The points within the microcanonical energy window $(\overline{E}-\sigma_{E,\text{q}},\overline{E}+\sigma_{E,\text{q}})$ are shown in red, while all other points are shown in blue.}
    \label{fig_C_a_IO_N_08}
\end{figure}

A plot of the coefficients $C_{\alpha}$ over the corresponding eigenstate energies $E_{\alpha}$ for $N=8$ is displayed in Fig.~\ref{fig_C_a_IO_N_08}.\footnote{In the chosen basis, all $C_{\alpha}$-s are real.} The points inside the energy window $(\overline{E}-\sigma_{E,\text{q}},\overline{E}+\sigma_{E,\text{q}})$ are shown in red, and the rest of the points are shown in blue.

We observe from Fig.~\ref{fig_C_a_IO_N_08} that the points closer to $\overline{E}=0$, in particular within the energy window, have greater absolute values of the coefficients $C_{\alpha}$. We find similar behavior for other values of $N$. Also recall from section~\ref{subsec_therm_test_num_an} that $\sigma_{E,\text{q}}$, compared to the ``shifted'' system energy $N$, is algebraically small in $N$. This suggests that the initial state (\ref{eqn_in_st}) is indeed a superposition of eigenstates that are sufficiently close in energy. We therefore find indications that the ETH condition (0) is satisfied, in particular in the TDL.

\subsection{Condition (1)}

We now test the ETH condition (1) in application to the observable $\hat{n}_1$. To satisfy the ETH condition (1), the diagonal matrix elements $n_{1,\alpha\alpha}$ must depend sufficiently smoothly on $E_{\alpha}$, and the absolute difference between neighboring elements $| n_{1;\alpha+1,\alpha+1} - n_{1;\alpha,\alpha} |$ must be exponentially small in $N$. We test both of these criteria below.

\subsubsection{Studied quantities} \label{subsubsec_ETH_test_1_qnt}

As an additional quantity, we first define the average of all diagonal matrix elements,
\begin{equation}
n_{i,\text{av}} \equiv \dfrac{1}{\mathcal{N}} \sum\limits_{\alpha=1}^{\mathcal{N}} n_{i,\alpha \alpha} \, .
\end{equation}
We can now appropriately normalize and compare the absolute differences $| n_{i;\alpha+1,\alpha+1} - n_{i;\alpha,\alpha} |$ for different $N$. For this purpose we introduce the normalized average absolute difference between neighboring diagonal matrix elements
\begin{equation}
\delta_i \equiv \dfrac{1}{n_{i,\text{av}}} \dfrac{1}{\mathcal{N}-1} \sum\limits_{\alpha=1}^{\mathcal{N}-1} |n_{i;\alpha+1,\alpha+1} - n_{i;\alpha,\alpha} | \, ,
\end{equation}
as well as its normalized corrected standard deviation
\begin{equation}
\sigma_i \equiv \dfrac{1}{n_{i,\text{av}}} \left[ \dfrac{1}{\mathcal{N}-2} \sum\limits_{\alpha=1}^{\mathcal{N}-1} \left( |n_{i;\alpha+1,\alpha+1} - n_{i;\alpha,\alpha} | - n_{i,\text{av}} \delta_i \right)^2 \right]^{1/2} \, .
\end{equation}

Analogously, we define corresponding quantities for the points within the microcanonical energy window,

\begin{equation}
\delta_{i,\text{mc}} \equiv \dfrac{1}{n_{i,\text{mc}}} \dfrac{1}{\mathcal{N}_{\sigma_{E,\text{q}}}-1} \sum\limits_{\substack{\alpha \\ |\overline{E} - E_{\alpha}| < \sigma_{E,\text{q}} \\ |\overline{E} - E_{\alpha+1}| < \sigma_{E,\text{q}}}} |n_{i;\alpha+1,\alpha+1} - n_{i;\alpha,\alpha} |
\end{equation}
and
\begin{equation}
\sigma_{i,\text{mc}} \equiv \dfrac{1}{n_{i,\text{mc}}} \left[ \dfrac{1}{\mathcal{N}_{\sigma_{E,\text{q}}}-2} \sum\limits_{\substack{\alpha \\ |\overline{E} - E_{\alpha}| < \sigma_{E,\text{q}} \\ |\overline{E} - E_{\alpha+1}| < \sigma_{E,\text{q}}}} \left( |n_{i;\alpha+1,\alpha+1} - n_{i;\alpha,\alpha} | - n_{i,\text{mc}} \delta_{i,\text{mc}} \right)^2 \right]^{1/2} \, .
\end{equation}

We also define the normalized maximum absolute difference among all pairs of neighboring diagonal matrix elements as
\begin{equation} \label{eqn_ETH_1_dlt_i_max}
\delta_{i,\text{max}} \equiv \dfrac{1}{n_{i,\text{av}}} \max \{ | n_{i;\alpha+1,\alpha+1} - n_{i;\alpha,\alpha} | \} \, .
\end{equation}

Last, we define the corresponding quantity for the points within the interval $(\overline{E}-\sigma_{E,\text{q}},\overline{E}+\sigma_{E,\text{q}})$, i.e.\ for $|\overline{E} - E_{\alpha}| < \sigma_{E,\text{q}}$ and $|\overline{E} - E_{\alpha+1}| < \sigma_{E,\text{q}}$, by
\begin{equation} \label{eqn_ETH_1_dlt_i_max_mc}
\delta_{\substack{i,\text{max} \\ \text{mc}}} \equiv \dfrac{1}{n_{i,\text{mc}}} \max_{\substack{|\overline{E} - E_{\alpha}| < \sigma_{E,\text{q}} \\ |\overline{E} - E_{\alpha+1}| < \sigma_{E,\text{q}}}} \{ | n_{i;\alpha+1,\alpha+1} - n_{i;\alpha,\alpha} | \} \, .
\end{equation}

\subsubsection{Numerical analysis}

\begin{figure}[htbp]
    \centering
    \begin{subfigure}[b]{0.45\textwidth}
        \caption{\footnotesize $n_{1,\alpha\alpha}$ over $E_{\alpha}$ for $N=8$}
        \includegraphics[width=\textwidth]{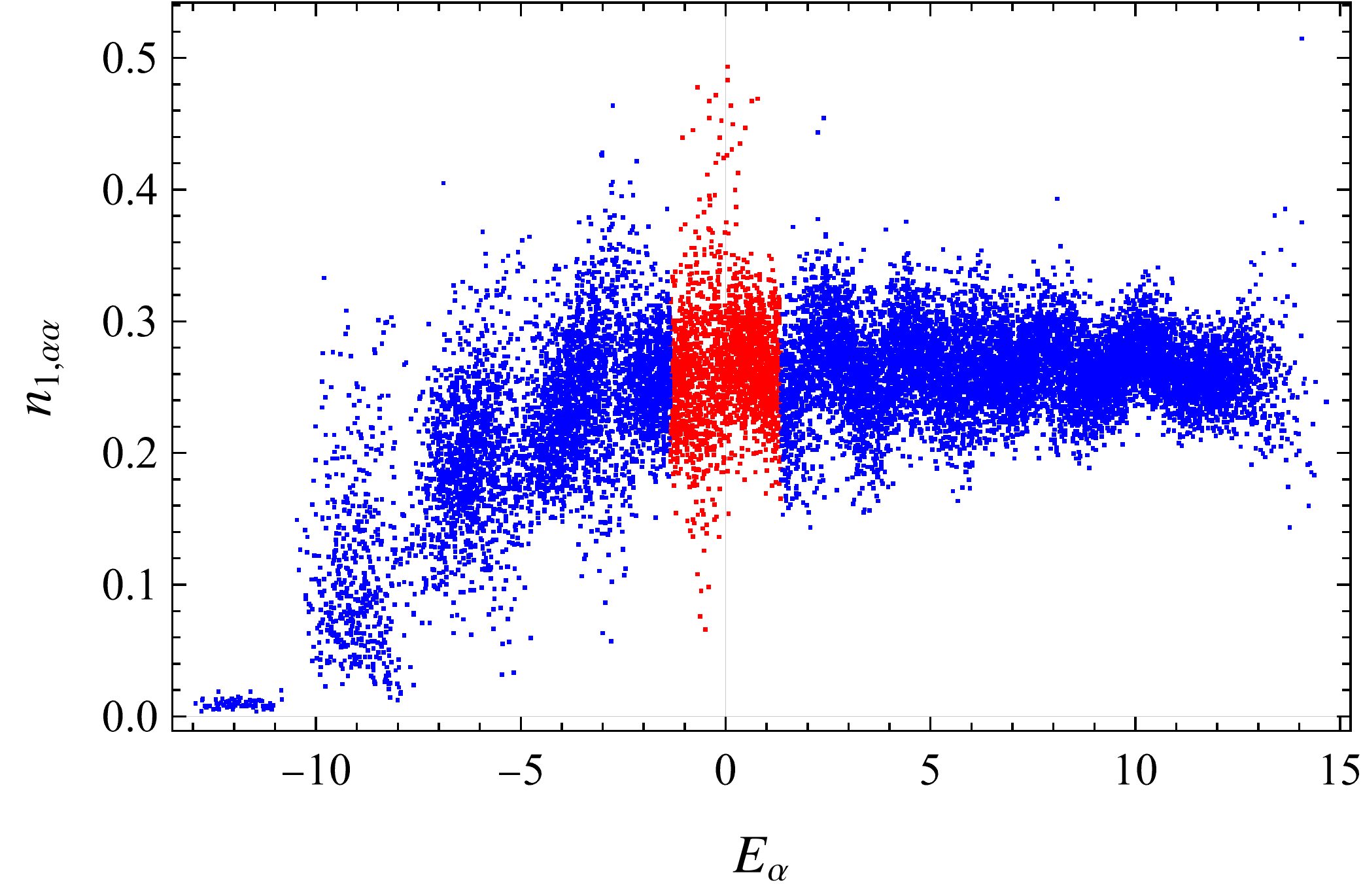}
        \centering
        \label{fig_n_0_aa_IO_N_08}
    \end{subfigure}
    \begin{subfigure}[b]{0.45\textwidth}
    	\caption{\footnotesize $n_{1,\text{av}} \sim (-0.25 \pm 2.6 \times 10^{-15}) N^{-1}$}
        \includegraphics[width=\textwidth]{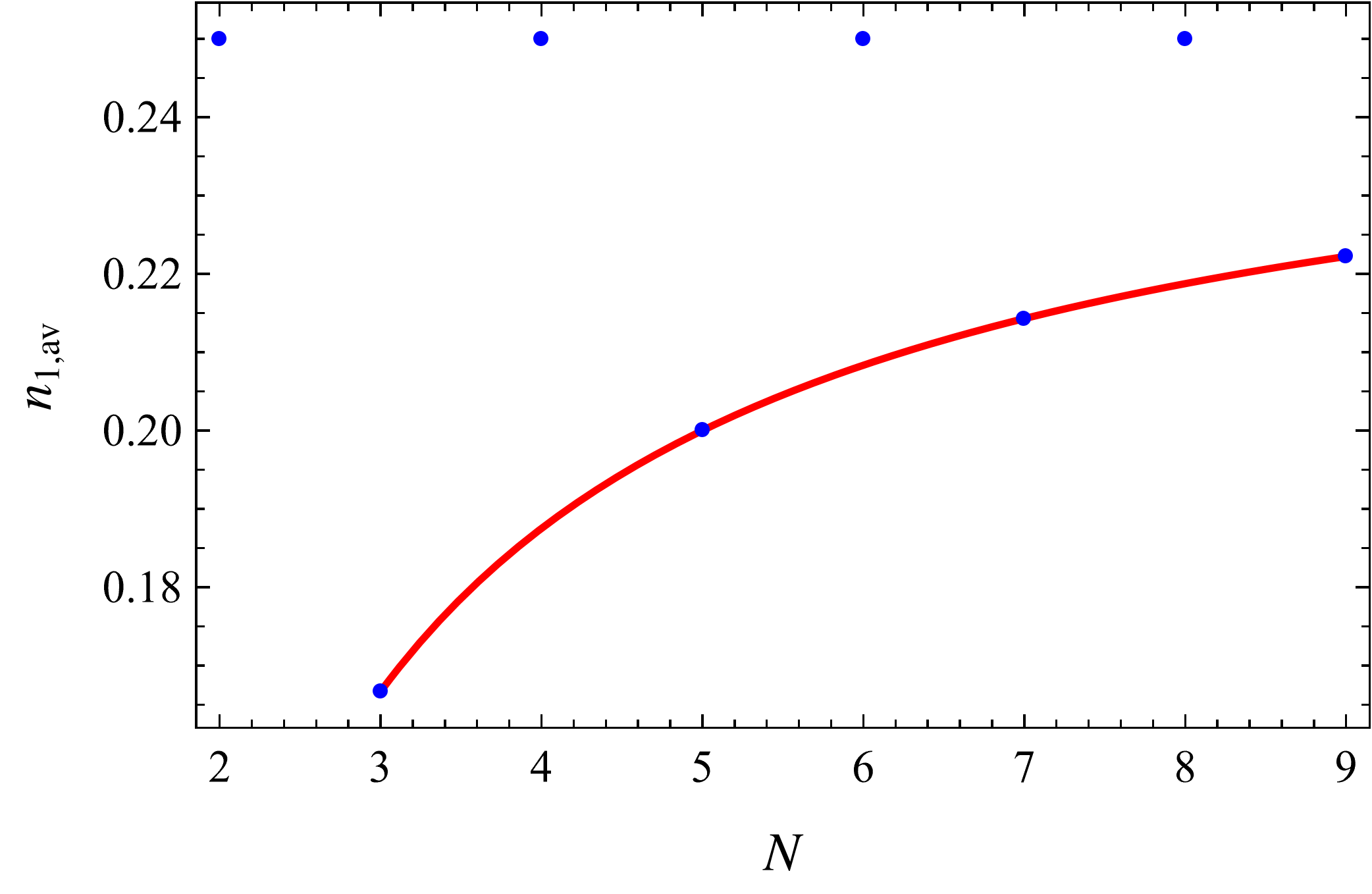}
        \label{fig_n0av_vs_N}
    \end{subfigure}
    \newline
    \begin{subfigure}[b]{0.32\textwidth}
    	\caption{\footnotesize $\delta_1 \sim \mathrm{e}^{(-0.29 \pm 0.66) N}$}
        \includegraphics[width=\textwidth]{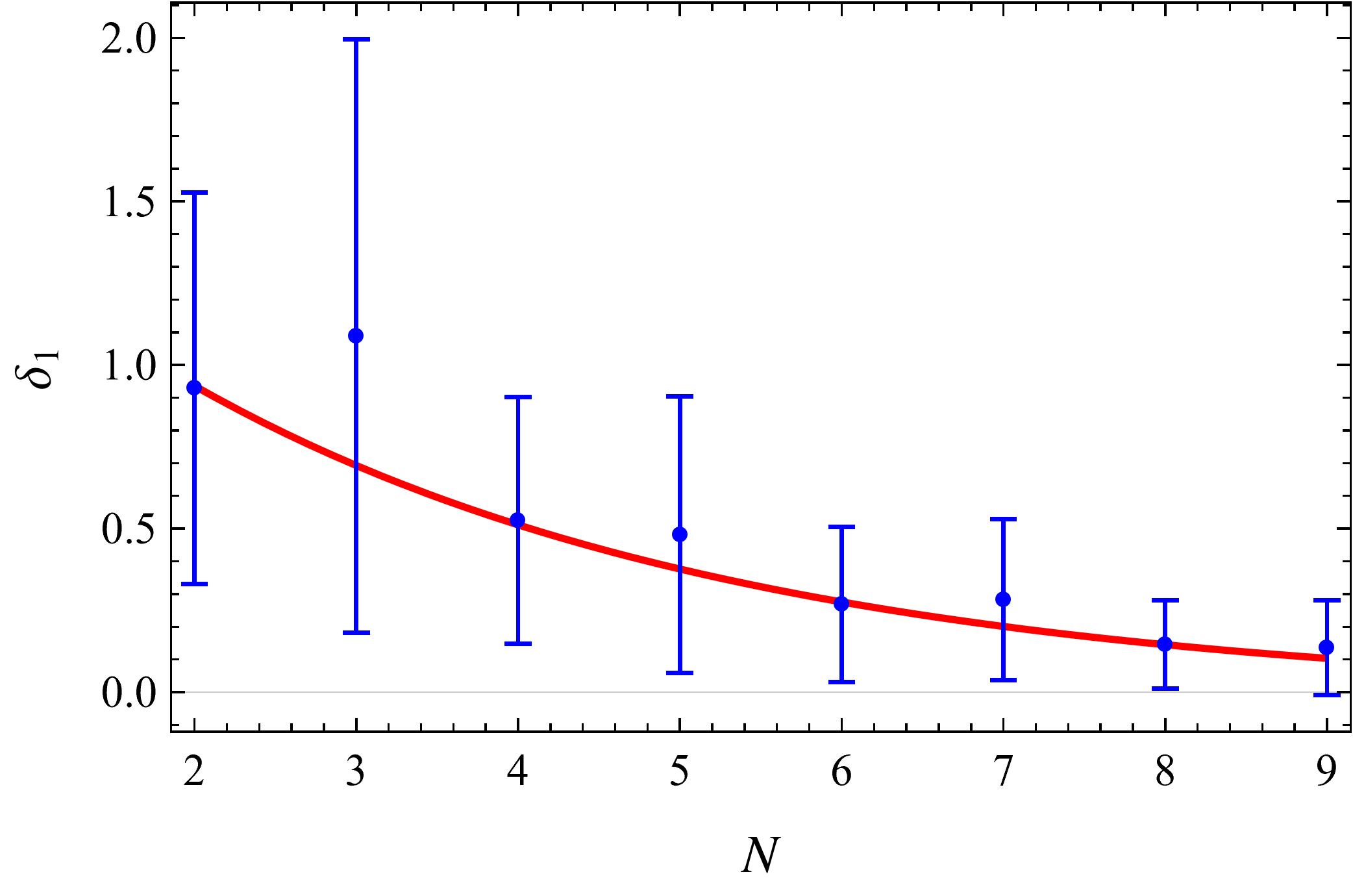}
        \label{fig_dlt0_vs_N}
    \end{subfigure}
    \begin{subfigure}[b]{0.32\textwidth}
    	\caption{\footnotesize $\delta_{1,\text{mc}} \sim -\mathrm{e}^{(-0.10 \pm 0.89) N}$}
        \includegraphics[width=\textwidth]{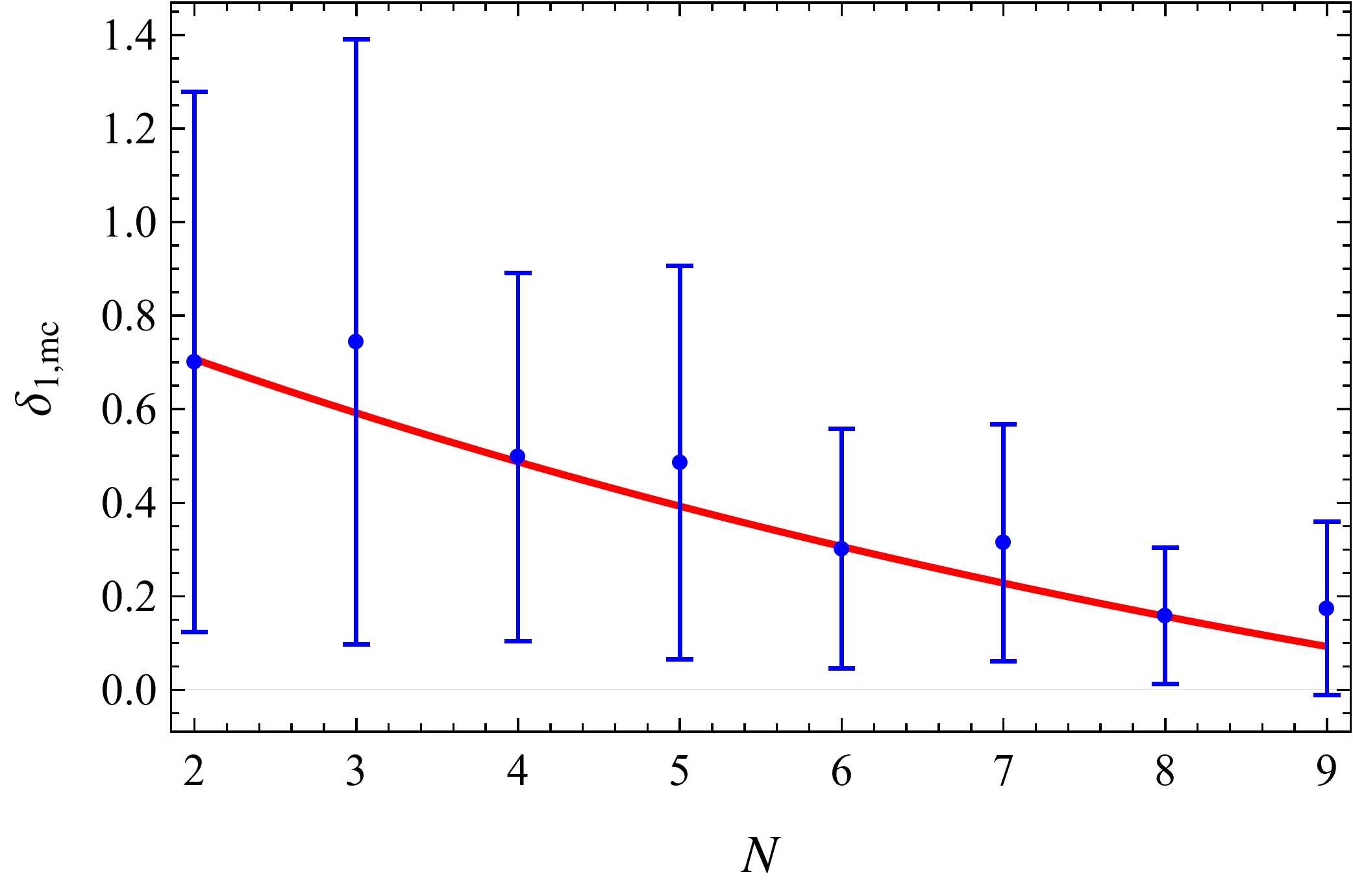}
        \label{fig_dlt0mc_vs_N}
    \end{subfigure}
        \begin{subfigure}[b]{0.32\textwidth}
    	\caption{\footnotesize $\delta_{\substack{1,\text{max} \\ \text{mc}}} \sim N^{-1.451 \pm 0.055}$}
        \includegraphics[width=\textwidth]{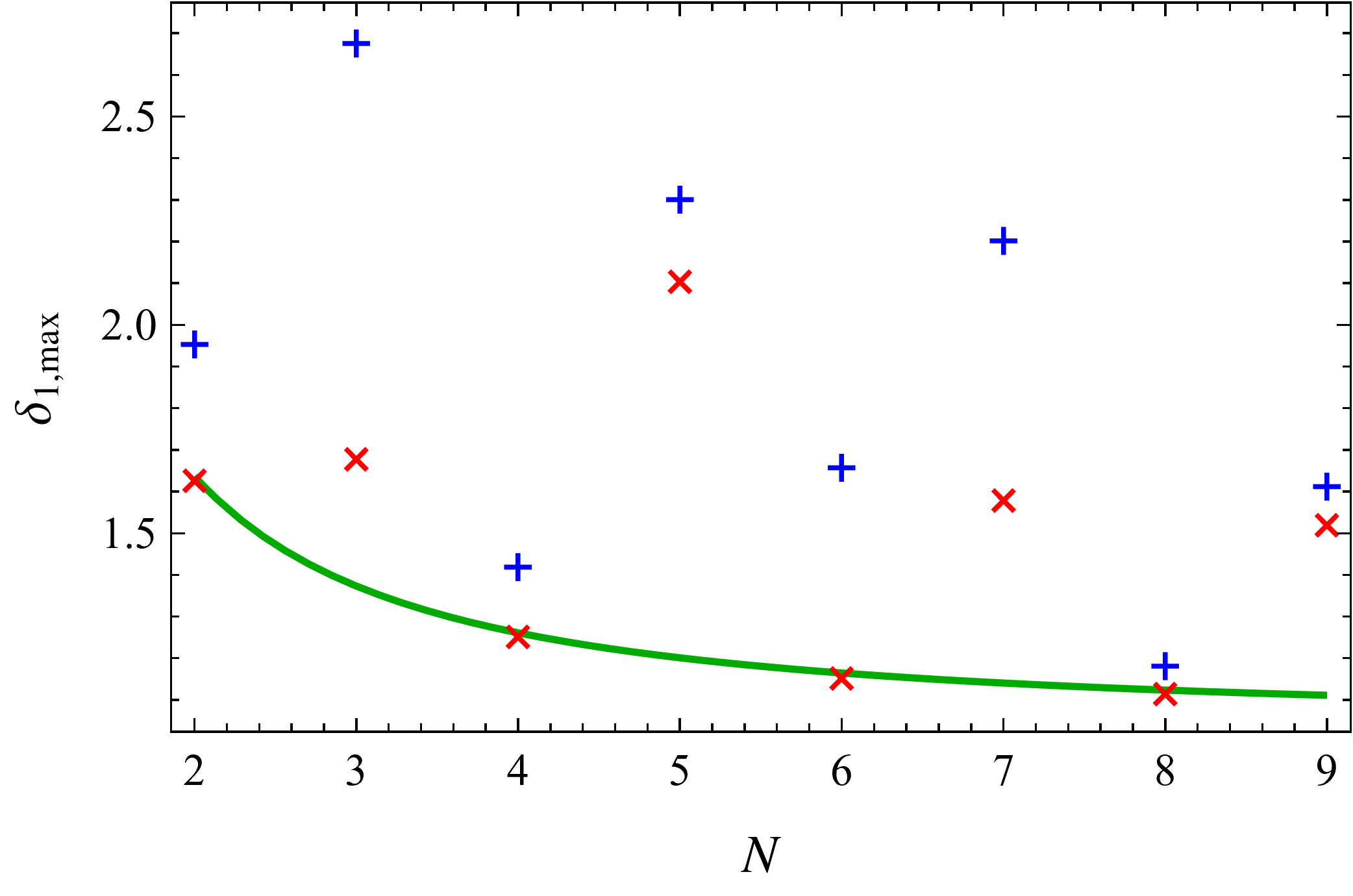}
        \label{fig_dlt0max_vs_N}
    \end{subfigure}
    \caption{\small Test of the ETH condition (1). Subfigure (\subref{fig_n_0_aa_IO_N_08}) displays a plot of $n_{1,\alpha \alpha}$ over $E_{\alpha}$ for $N=8$. The points within the microcanonical energy shell $\overline{E} \pm \sigma_{E,\text{q}}$ are shown in red, all other points are shown in blue. The other subfigures display the quantities from section~\ref{subsubsec_ETH_test_1_qnt} plotted over $N$. As before, the subcaptions show only the $N$-scaling. The fit parameter values are given in Table~\ref{tab_fit} in the Appendix. The numerical data is shown with points. The respective fits are represented as solid lines. The subfigures display: (\subref{fig_n0av_vs_N}) $n_{1,\text{av}}$ where the respective fit is performed only over the points with odd $N$, (\subref{fig_dlt0_vs_N}) $\delta_1$ with $\sigma_1$ represented by error bars, (\subref{fig_dlt0mc_vs_N}) $\delta_{1,\text{mc}}$ with $\sigma_{1,\text{mc}}$ as error bars, (\subref{fig_dlt0max_vs_N}) $\delta_{1,\text{max}}$ shown as blue ``$+$''-s, $\delta_{1,\text{max}; \text{mc}}$ shown as red ``$\times$''-s and the fit of the latter points with even $N$ shown as a solid green line. The fits of $\delta_1$ and $\delta_{1,\text{mc}}$ are performed only over the points with even $N$ and are weighted, with the weights $\sigma_1^{-2}$ and $\sigma_{1,\text{mc}}^{-2}$ for the corresponding data, respectively.}
    \label{fig_ETH_1}
\end{figure}

Figure~\ref{fig_ETH_1} shows the numerical results for the quantities in section~\ref{subsubsec_ETH_test_1_qnt}. We observe from an exemplary plot for $N=8$ in Fig.~\ref{fig_n_0_aa_IO_N_08} that the diagonal elements $n_{1,\alpha \alpha}$ do not vary smoothly with $E_{\alpha}$. We find similar behavior in plots of $n_{1,\alpha \alpha}$ over $E_{\alpha}$ for other $N$. This already indicates a disagreement with the first part of the ETH condition (1) in section~\ref{subsubsec_ETH_cond}. However, the crucial behavior of the system is that in the TDL. We therefore test whether the absolute differences between neighboring diagonal matrix elements $| n_{1;\alpha+1,\alpha+1} - n_{1;\alpha,\alpha} |$ are exponentially small in $N$.

As a preliminary step, we first observe from Fig.~\ref{fig_n0av_vs_N} that $n_{1,\text{av}} = 0.25$ holds for even values of $N$. From the fit parameter values in Table~\ref{tab_fit} we find that for odd $N$ the value of $n_{1,\text{av}}$ approaches $0.25$ as $\mathcal{O}(N^{-1})$. To reduce the effects of small values of $N$, we use only the points with even $N$ for the fits of $\delta_1$ and $\delta_{1,\text{mc}}$. Figures~\ref{fig_dlt0_vs_N} and~\ref{fig_dlt0mc_vs_N} display the corresponding results, respectively. From the fit parameter values in Table~\ref{tab_fit} we find that the fits of both $\delta_1$ and $\delta_{1,\text{mc}}$ are consistent with a vanishing value in the large-$N$ limit.

We can explain this as follows: First, for all considered $N$, we observe that the majority of the diagonal matrix elements $n_{1,\alpha \alpha}$ are clustered around $n_{1,\text{av}}$. Second, both their total number $\mathcal{N}$ and their number within the microcanonical energy shell $\mathcal{N}_{\sigma_{E,\text{q}}}$ increase exponentially with $N$ (for $\mathcal{N}_{\sigma_{E,\text{q}}}$, see Fig.~\ref{fig_NsigEq_vs_N} and the corresponding fit in Table~\ref{tab_fit}). Therefore, the absolute difference between most neighboring diagonal elements decreases with $N$. Specifically, we conclude that \emph{on average} the absolute differences $| n_{1;\alpha+1,\alpha+1} - n_{1;\alpha,\alpha} |$ are exponentially small in $N$ (see fits of $\delta_1$ and $\delta_{1,\text{mc}}$ in Table~\ref{tab_fit}). Nevertheless, the differences for particular pairs of diagonal elements can still be significant.

To test this, we examine the normalized maximum absolute difference $\delta_{1,\text{max}}$ in (\ref{eqn_ETH_1_dlt_i_max}) between pairs of neighboring diagonal elements for all $\mathcal{N}$ eigenstates of the spectrum, as well as the analogous $\delta_{1,\text{max}; \text{mc}}$ in (\ref{eqn_ETH_1_dlt_i_max_mc}) for the $\mathcal{N}_{\sigma_{E,\text{q}}}$ eigenstates $\ket{\alpha}$ with $E_{\alpha} \in (\overline{E}-\sigma_{E,\text{q}},\overline{E}+\sigma_{E,\text{q}})$. Figure~\ref{fig_dlt0max_vs_N} displays the data for $\delta_{1,\text{max}}$ and $\delta_{1,\text{max}; \text{mc}}$ in blue and red, respectively. To avoid relying on properties of eigenstates that lie far outside the bulk of the spectrum, we base our argument on the eigenstates within the microcanonical energy window and the corresponding quantity $\delta_{1,\text{max}; \text{mc}}$. For the fit of $\delta_{1,\text{max}; \text{mc}}$ we consider only the data points with even $N$ to minimize small-$N$ effects.

We observe from Fig.~\ref{fig_dlt0max_vs_N} that the normalized maximum absolute differences are the smallest for even $N$ and for eigenstates $\ket{\alpha}$ with $E_{\alpha} \in (\overline{E}-\sigma_{E,\text{q}},\overline{E}+\sigma_{E,\text{q}})$. The best fit of $\delta_{1,\text{max}; \text{mc}}$ in Table~\ref{tab_fit} suggests that in the TDL it approaches a value within the interval $[1.0350,1.0534]$. That is, we find indications that in the large-$N$ limit the maximum absolute difference between neighboring diagonal elements within the microcanonical energy window exceeds the microcanonical average itself. This is a substantial variation.

To summarize, we find indications that the ETH condition (1) is not satisfied. Specifically, we observe that the diagonal elements $n_{1,\alpha\alpha}$ do not vary smoothly with $E_{\alpha}$ and that the magnitude of the maximal difference between neighboring diagonal elements is not exponentially small in $N$. The numerical results suggest that this behavior persists in the large-$N$ limit.

However, the analysis of section~\ref{subsec_therm_test_num_an} provides numerical evidence that the observable $\hat{n}_1$ thermalizes in the TDL. In sections~\ref{sec_TDC} and \ref{sec_disc} we discuss the corresponding proposed thermalization mechanism.

\subsection{Condition (2)}

In this section we test the ETH condition (2) in application to the observable $\hat{n}_1$. To satisfy the ETH condition (2), the absolute values of the off-diagonal matrix elements $|n_{1,\alpha \beta; \alpha \neq \beta}|$ must be exponentially small in $N$.

The properties of the off-diagonal elements are directly related to the temporal behavior of the system and therefore also to its thermalization: In a system with a non-degenerate spectrum, the time-dependence of the expectation value $\braket{\hat{A}(t)}$ of an observable $\hat{A}$ is controlled by the off-diagonal matrix elements $A_{\alpha \beta}$, along with the coefficients $C_{\alpha}$ and $C_{\beta}$, and eigenenergies $E_{\alpha}$ and $E_{\beta}$, for $\alpha \neq \beta$. Specifically,
\begin{equation}
\braket{\hat{A}(t)} = \sum\limits_{\alpha, \beta} C_{\alpha}^* C_{\beta} A_{\alpha \beta} \mathrm{e}^{i (E_{\alpha}-E_{\beta}) t} \, .
\end{equation}

Recall also from (\ref{eqn_temp_fluct_sigma_i_t}) that the infinite-time root mean squared magnitude of the temporal fluctuations of the observable $\hat{A}$ about the infinite-time average $\overline{A}$ of its expectation value $\braket{\hat{A}(t)}$ is given by
\begin{equation}
\sigma_t = \left[ \overline{\braket{\hat{A}(t)}^2}-\overline{A}^2 \right]^{1/2} = \left[ \sum\limits_{\substack{\alpha, \beta \\ \alpha \neq \beta}} |C_{\alpha}|^2 |C_{\beta}|^2 |A_{\alpha\beta}|^2 \right]^{1/2} \, .
\end{equation}

\begin{figure}[htbp]
    \centering
    \begin{subfigure}[c]{0.45\textwidth}
        \caption{\footnotesize $|n_{1, \alpha \beta}|$ over $\alpha$ and $\beta$ for $N=4$}
        \includegraphics[width=\textwidth]{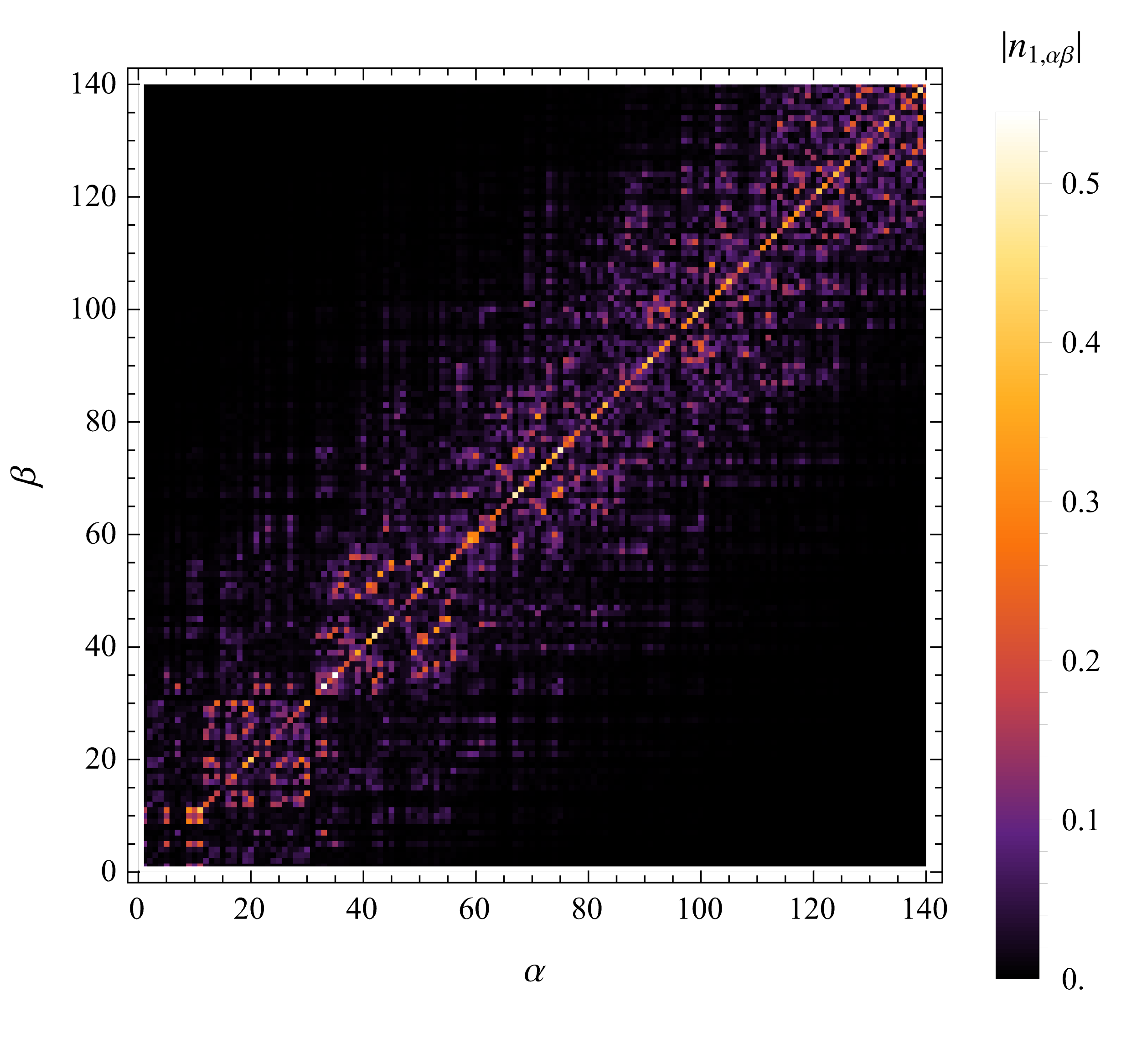}
        \centering
        \label{fig_abs_n_0_ab_N_04}
    \end{subfigure}
    \begin{subfigure}[c]{0.45\textwidth}
    	\caption{\footnotesize $|n_{1,\alpha \beta ; \alpha \neq \beta}|_{\text{av}} \sim \mathrm{e}^{(-0.654 \pm 0.082) N}$}
        \includegraphics[width=\textwidth]{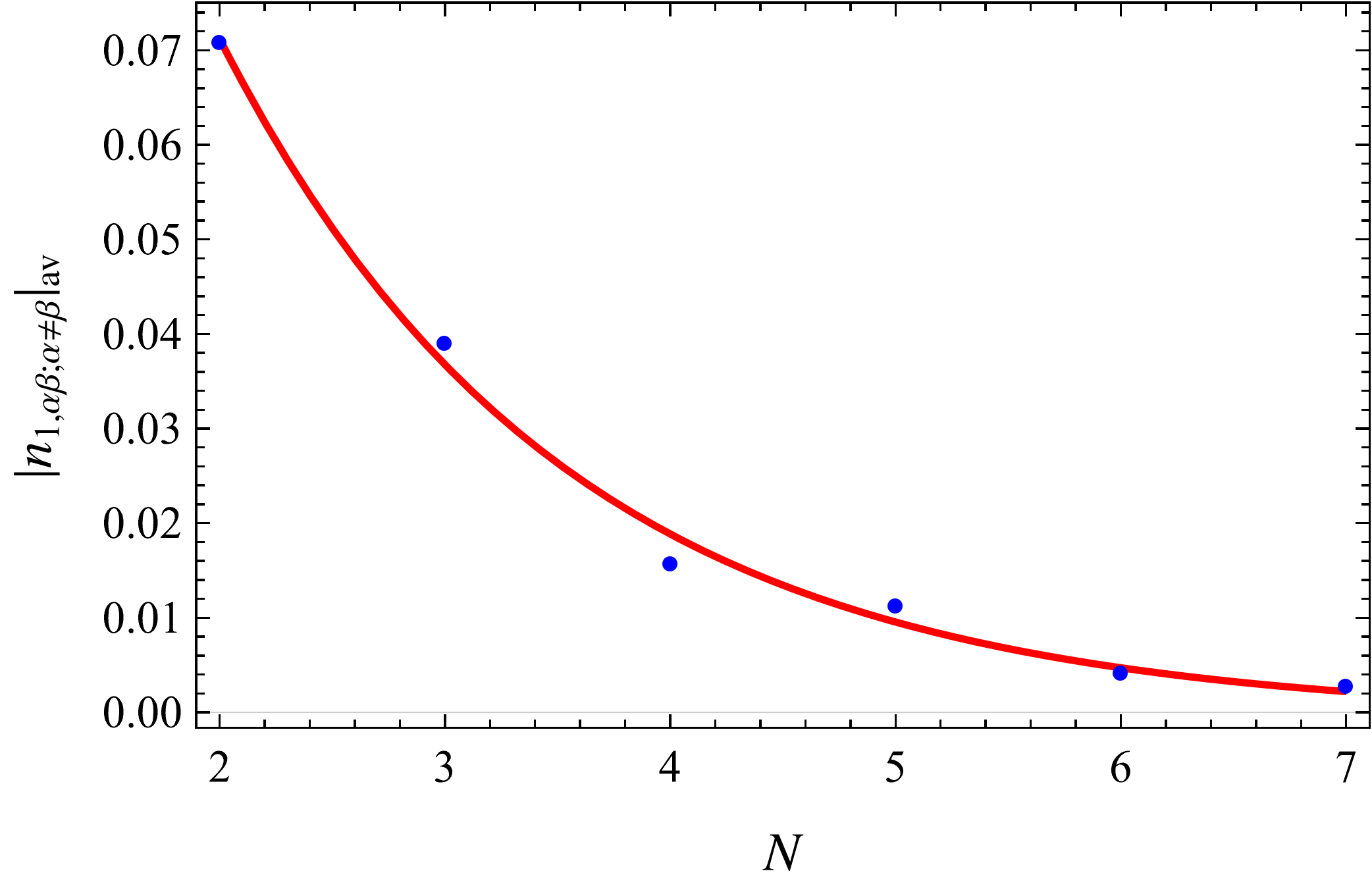}
        \label{fig_av_abs_n0ab_vs_N}
    \end{subfigure}
    \caption{\small Test of the ETH condition (2). Subfigure (\subref{fig_abs_n_0_ab_N_04}) displays a plot of $|n_{1,\alpha\beta}|$ over $\alpha$ and $\beta$ for $N=4$. Subfigure (\subref{fig_av_abs_n0ab_vs_N}) shows the mean absolute value $|n_{1,\alpha \beta ; \alpha \neq \beta}|_{\text{av}}$ of the off-diagonal matrix elements over $N$. The data is represented by points. The respective fit is shown as a solid line. See Table~\ref{tab_fit} in the Appendix for the respective fit parameter values.}
    \label{fig_ETH_2}
\end{figure}

In section~\ref{subsec_therm_test_num_an}, we provided evidence that the thermalization condition (ii) on the temporal fluctuations of the observable $\hat{n}_1$ is fulfilled in the large-$N$ limit. In this section we argue that the magnitudes of the off-diagonal matrix elements $|n_{1,\alpha \beta; \alpha \neq \beta}|$ satisfy the ETH condition (2). Figure~\ref{fig_ETH_2} shows the corresponding numerical results. We observe from Fig.~\ref{fig_abs_n_0_ab_N_04} for $N=4$ that the magnitudes of the diagonal elements are generally larger than the magnitudes of the off-diagonal ones. We find similar behavior in plots of $|n_{1,\alpha\beta}|$ over $\alpha$ and $\beta$ for other $N$.

We also consider how the average absolute value of the off-diagonal matrix elements
\begin{equation}
|n_{i,\alpha \beta ; \alpha \neq \beta}|_{\text{av}} \equiv \dfrac{1}{\mathcal{N}(\mathcal{N}-1)} \sum\limits_{\substack{\alpha, \beta \\ \alpha \neq \beta}} |n_{i,\alpha \beta}|
\end{equation}
for the observable $\hat{n}_1$ depends on $N$. The results are shown in Fig.~\ref{fig_av_abs_n0ab_vs_N}. From the corresponding fit in Table~\ref{tab_fit} we find evidence that $|n_{1,\alpha \beta ; \alpha \neq \beta}|_{\text{av}}$ is exponentially small in $N$.

\section{Tests of other mechanisms} \label{sec_test_other_mech}

In section~\ref{sec_therm_test} we provided evidence that the observable $\hat{n}_1$ thermalizes in the large-$N$ limit. In section~\ref{sec_ETH_test} we argued that the ETH is not the underlying mechanism for this thermalization, as the diagonal matrix elements $n_{1, \alpha \alpha}$ do not satisfy the ETH condition (1). In this section and in section~\ref{sec_TDC} we justify why other thermalization mechanisms, specifically~\cite{Berges_2004, Rigol_2008, Biroli_2010, Ikeda_2011, Anza_2018, Li_2018}, cannot explain the thermalization of the observable $\hat{n}_1$.

The coefficients $C_{\alpha}$ fluctuate considerably between eigenstates close in energy, particularly within the microcanonical energy window (see Fig.~\ref{fig_C_a_IO_N_08}). The same holds for the squared magnitudes of the coefficients $|C_{\alpha}|^2$. Consequently, the condition of the mechanism (ii) of~\cite{Rigol_2008} (see section~\ref{subsubsec_Rigol_mech}) is not fulfilled.

The fit results of $\delta_{1,\text{max}; \text{mc}}$ in Table~\ref{tab_fit} in the Appendix (see also Fig.~\ref{fig_dlt0max_vs_N}) suggest that the fluctuations of the diagonal matrix elements $n_{1,\alpha\alpha}$ are non-zero in the large-$N$ limit. Therefore, the conditions for the \emph{thermal} and the \emph{smoothness} variants~\cite{Anza_2018} of the ETH are also not satisfied.

The thermalization of $\hat{n}_1$ also cannot occur via \emph{thermalization due to integrability}~\cite{Li_2018}: We regard the model (\ref{eqn_model}) as non-integrable. Specifically, from the viewpoint of~\cite{Li_2018}, we consider it to be not analytically solvable. For non-integrability from the perspective of level spacing statistics, recall the discussion on the model's level spacing distribution in section~\ref{subsubsec_model}.

Since we consider the model (\ref{eqn_model}) to be non-integrable, we also argue that it is not nearly integrable. In addition, we do not find two distinct relaxation time scales. Specifically, we do not observe equilibration to a meta-stable state on a short time scale followed by an approach towards the true thermal equilibrium on a longer time scale. Therefore, the observable $\hat{n}_1$ does not thermalize via the mechanism of \emph{prethermalization}~\cite{Berges_2004}.

We now address two variations of the ETH, namely the \emph{strong ETH} and the \emph{weak ETH}~\cite{Biroli_2010}. The determining feature for both of these mechanisms is the distribution of the diagonal elements. From the perspective of these two mechanisms, two aspects characterize this distribution: First, the fraction of eigenstates $\ket{\alpha}$ that result in a non-thermal expectation value of an observable $\hat{A}$, and second, the support of the distribution of the diagonal matrix elements $A_{\alpha \alpha}$, that is, the actual values of the quantities $A_{\alpha \alpha}$.

In the framework of the strong ETH, the rare non-thermal eigenstates disappear entirely in the TDL. Namely, the support of the distribution of the $A_{\alpha \alpha}$-s around the thermal value decreases to zero. In application to the observable $\hat{n}_1$ within the model (\ref{eqn_model}), the fit results of $\delta_{1,\text{max}; \text{mc}}$ in Table~\ref{tab_fit} in the Appendix (see also Fig.~\ref{fig_dlt0max_vs_N}) suggest that the non-thermal eigenstates persist in the large-$N$ limit. Specifically, we find that the maximum of $|n_{1;\alpha+1,\alpha+1}-n_{1;\alpha,\alpha}|$ for eigenstates $\ket{\alpha}$, $\ket{\alpha+1}$ within the microcanonical energy shell approaches a value of approximately $n_{1,\text{mc}}$.

Our results thus indicate that in the TDL, while $\overline{n}_1$ and $n_{1,\text{mc}}$ approach the thermal value of $0.25$, a non-thermal eigenstate $\ket{\alpha}$ exists within the microcanonical energy window with a corresponding value of $n_{1,\alpha \alpha}$ such that $|n_{1,\text{mc}}-n_{1,\alpha \alpha}| \gtrsim n_{1,\text{mc}}/2$. We conclude that this mechanism cannot explain the thermalization of the observable $\hat{n}_1$ within the model (\ref{eqn_model}).

Within the mechanism of the weak ETH, the fraction of the non-thermal states decreases to zero in the TDL. Therefore, the distribution of the $A_{\alpha \alpha}$-s shrinks around the thermal value. However, the support does not contract towards the thermal value. Namely, rare non-thermal eigenstates persist in the TDL, but the $|C_{\alpha}|^2$-s do not bias them sufficiently enough. Our results in section~\ref{sec_TDC} indicate that this thermalization mechanism is not applicable to the observable $\hat{n}_1$ within the model (\ref{eqn_model}).

In addition, let us remark that~\cite{Biroli_2010} also proposes that a plausible, but not necessary, assumption for the thermalization by the mechanism of the weak ETH is that the $|C_{\alpha}|^2$-s sample the eigenstates with the same energy rather uniformly. Note that other works, e.g.~\cite{Steinigeweg_2013, Beugeling_2014}, also suggest the randomness of the $|C_{\alpha}|^2$-s in application to the ETH.

The randomness of various quantities is an essential component of certain thermalization mechanisms. These include the mechanism (i) of~\cite{Rigol_2008} (see section~\ref{subsubsec_Rigol_mech}) and the \emph{Eigenstate Randomization Hypothesis} (ERH)~\cite{Ikeda_2011}. Within the mechanism (i) of~\cite{Rigol_2008}, the fluctuations of the diagonal elements $A_{\alpha \alpha}$ and those of the coefficients $|C_{\alpha}|^2$ are not correlated. In the framework of the ERH the $A_{\alpha \alpha}$-s fluctuate randomly.

In section~\ref{sec_TDC} we provide numerical evidence that the conditions for both of the above mechanisms are not satisfied for the observable $\hat{n}_1$ within the model (\ref{eqn_model}). Specifically, our results suggest that the fluctuations of the $n_{1, \alpha \alpha}$-s and those of the $|C_{\alpha}|^2$-s remain significantly correlated in the TDL.

\section{Thermalization despite correlation} \label{sec_TDC}

In this section we argue that neither the weak ETH, nor the mechanism (i) of~\cite{Rigol_2008} (see section~\ref{subsubsec_Rigol_mech}), nor the ERH are responsible for the thermalization of $\hat{n}_1$ within the model (\ref{eqn_model}). Our reasoning is based on a numerical observation indicating that the correlations in the fluctuations of the $C_{\alpha}^2$-s and the $n_{1,\alpha \alpha}$-s remain considerable in the limit of large $N$. We define the normalized fluctuations of the diagonal matrix elements $n_{i,\alpha\alpha}$ about their average $n_{i,\text{av}}$ as
\begin{equation}
\Delta_{n_{i,\alpha\alpha}} \equiv \dfrac{n_{i,\alpha\alpha}-n_{i,\text{av}}}{n_{i,\text{av}}} = \dfrac{n_{i,\alpha\alpha}}{n_{i,\text{av}}} - 1 \, .
\end{equation}
The normalized fluctuations of the coefficients-squared $C_{\alpha}^2$ are given by
\begin{equation}
\Delta_{C_{\alpha}^2} \equiv \dfrac{C_{\alpha}^2 - 1/\mathcal{N}}{1/\mathcal{N}} = C_{\alpha}^2 \mathcal{N} - 1 \, .
\end{equation}

\begin{figure}[htbp]
    \centering
    \begin{subfigure}[b]{0.32\textwidth}
        \caption{\footnotesize $\Delta_{n_{1,\alpha \alpha}}$ over $E_{\alpha}$}
        \includegraphics[width=\textwidth]{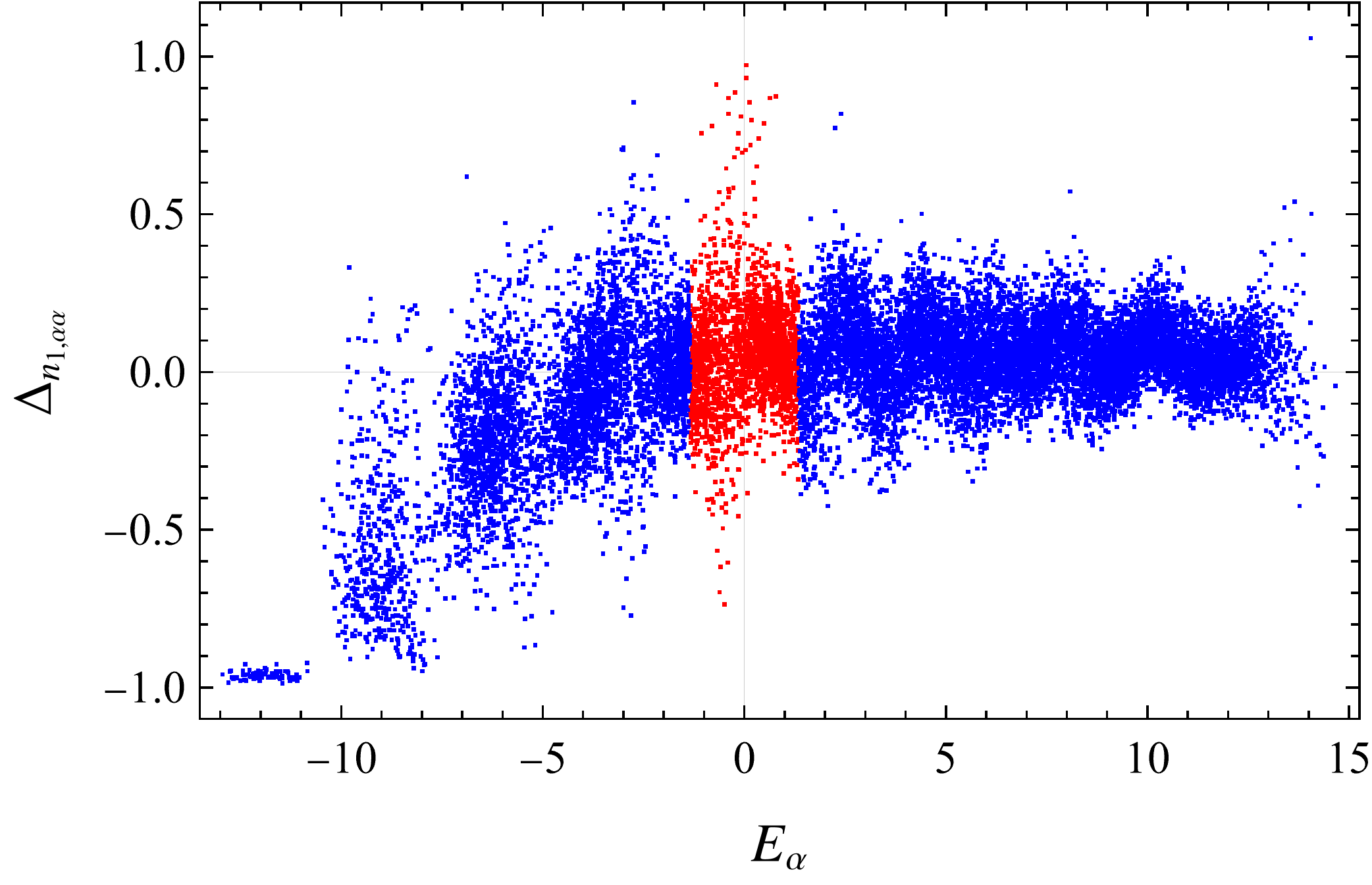}
        \centering
        \label{fig_DLT_n_0_aa_IO_N_08}
    \end{subfigure}
    \begin{subfigure}[b]{0.32\textwidth}
    	\caption{\footnotesize $\Delta_{C_{\alpha}^2}$ over $E_{\alpha}$}
        \includegraphics[width=\textwidth]{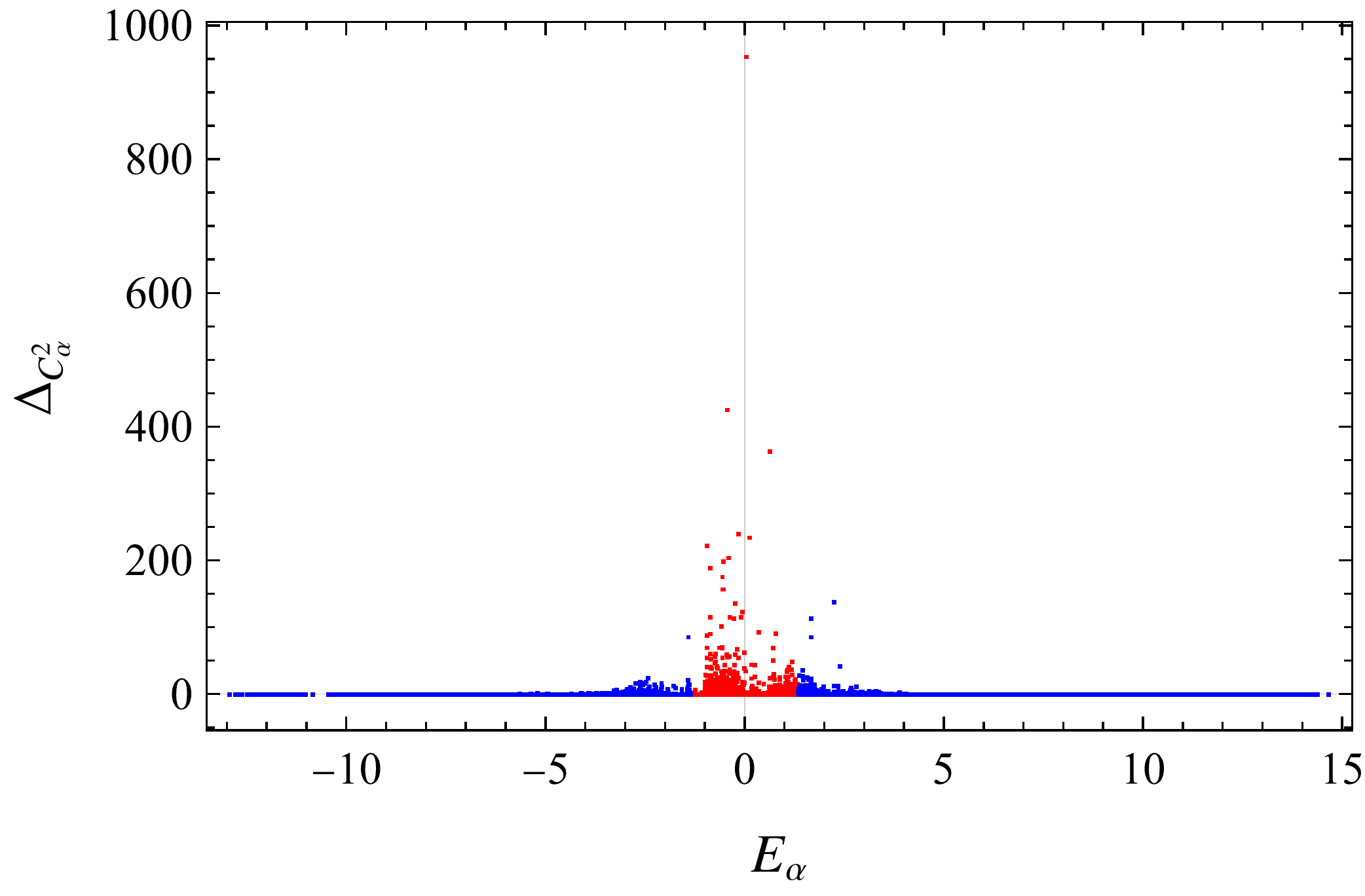}
        \label{fig_DLT_C_a_sq_IO_N_08}
    \end{subfigure}
    \begin{subfigure}[b]{0.32\textwidth}
    	\caption{\footnotesize $\Delta_{C_{\alpha}^2}$ over $\Delta_{n_{1,\alpha \alpha}}$}
        \includegraphics[width=\textwidth]{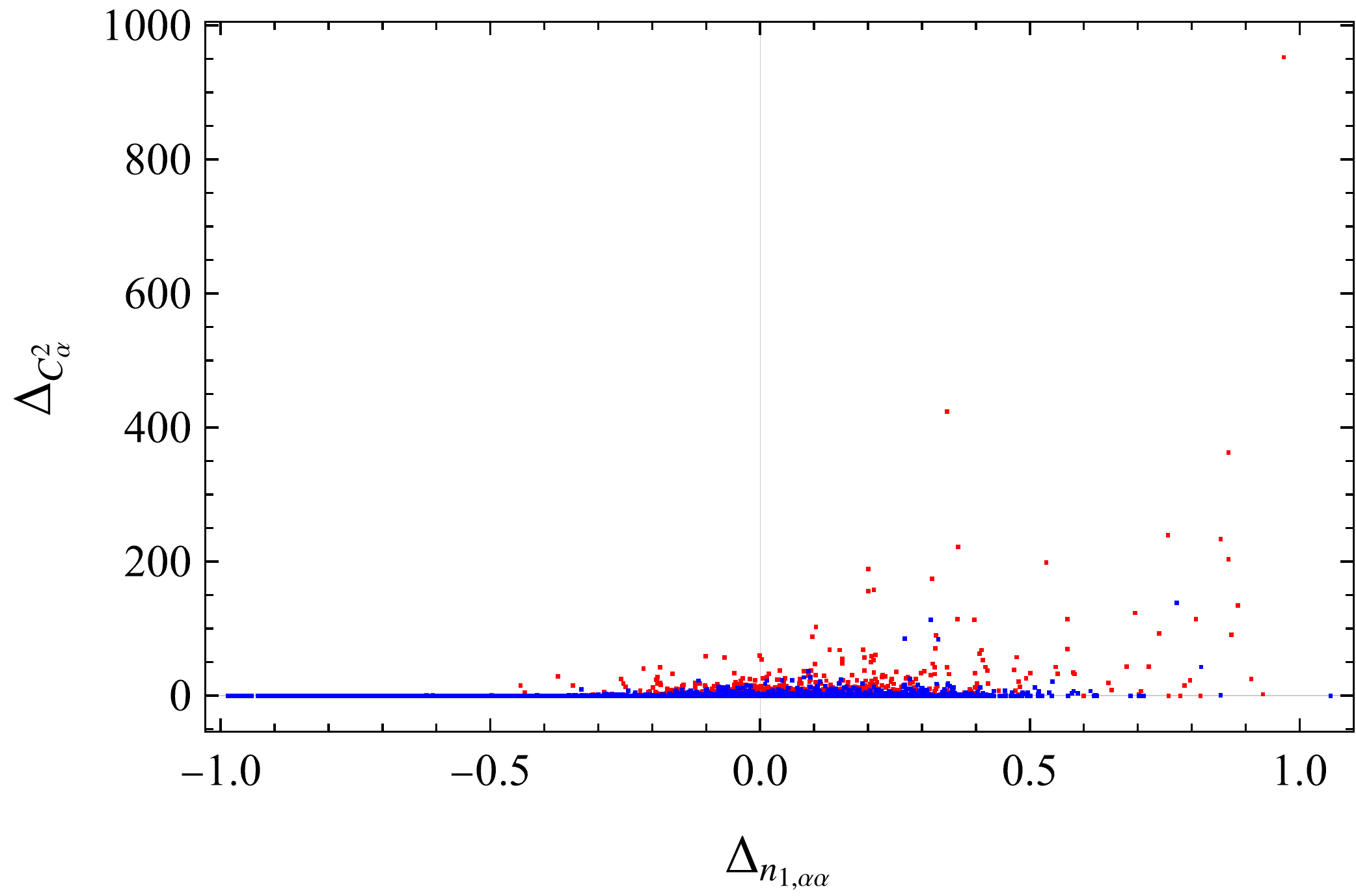}
        \label{fig_DLT_n_0_DLT_C_IO_N_08}
    \end{subfigure}
    \begin{subfigure}[b]{0.32\textwidth}
    	\caption{\footnotesize Count $c$ over $n_{1,\alpha \alpha}$ and $C_{\alpha}^2$}
    	\includegraphics[width=\textwidth]{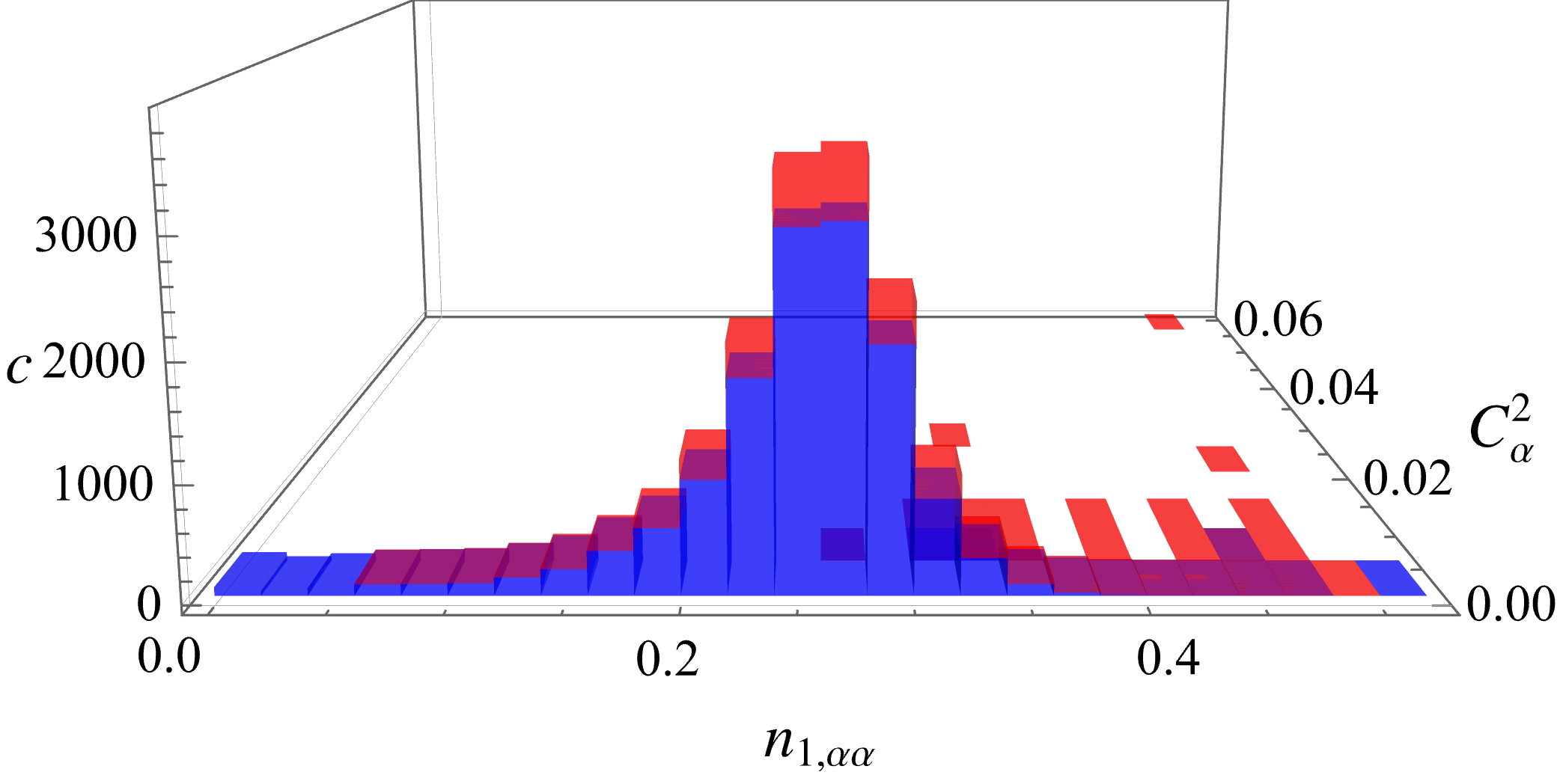}
        \label{fig_hist_N_08}
    \end{subfigure}
    \begin{subfigure}[b]{0.32\textwidth}
    	\caption{\footnotesize Count $c$ over $n_{1,\alpha \alpha}$}
    	\includegraphics[width=\textwidth]{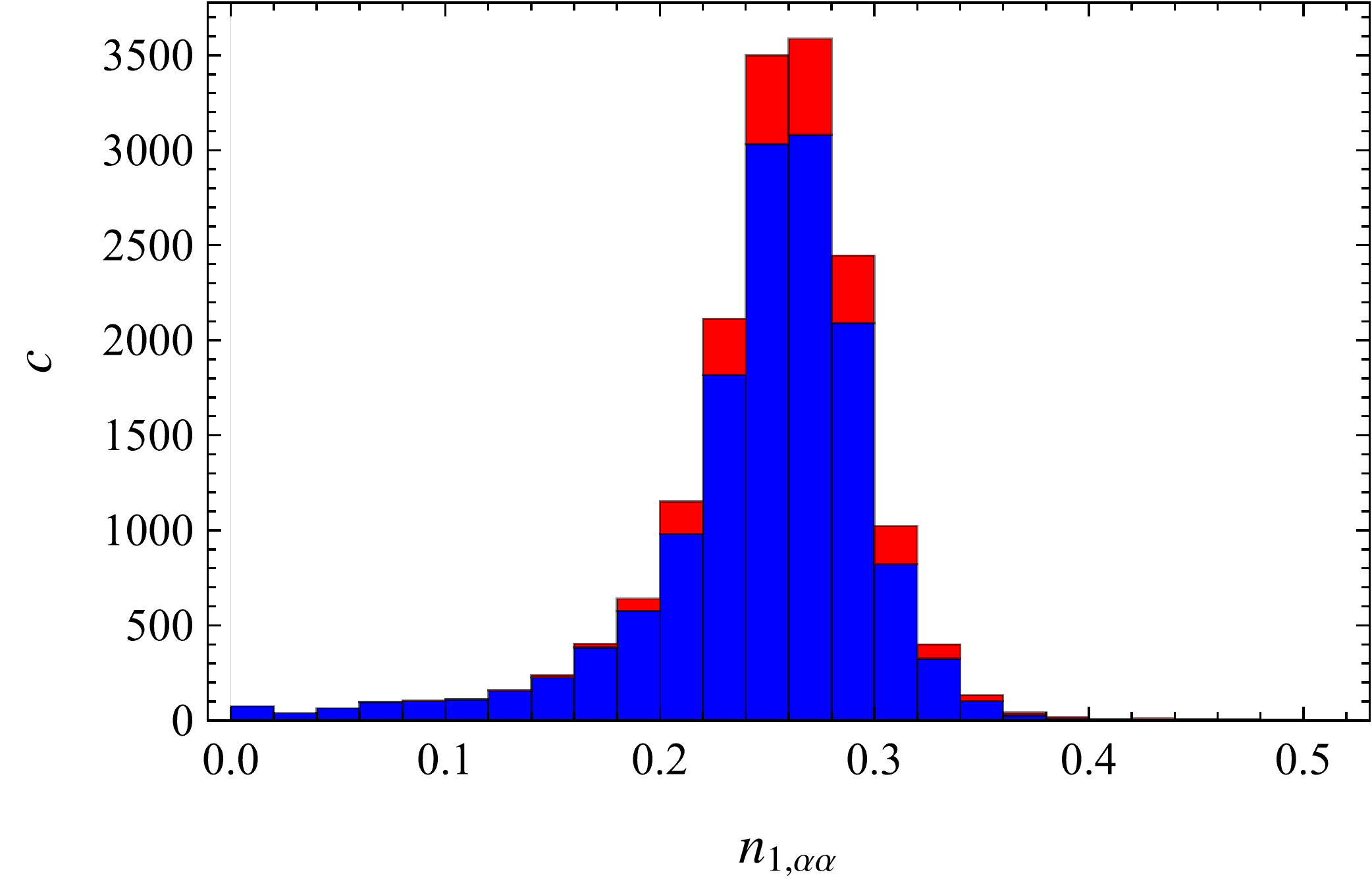}
        \label{fig_hist_ALL_N_08}
    \end{subfigure}
    \begin{subfigure}[b]{0.32\textwidth}
    	\caption{\footnotesize Count $c$ over $n_{1,\alpha \alpha}$ in $\sigma_{E,\text{q}}$}
    	\includegraphics[width=\textwidth]{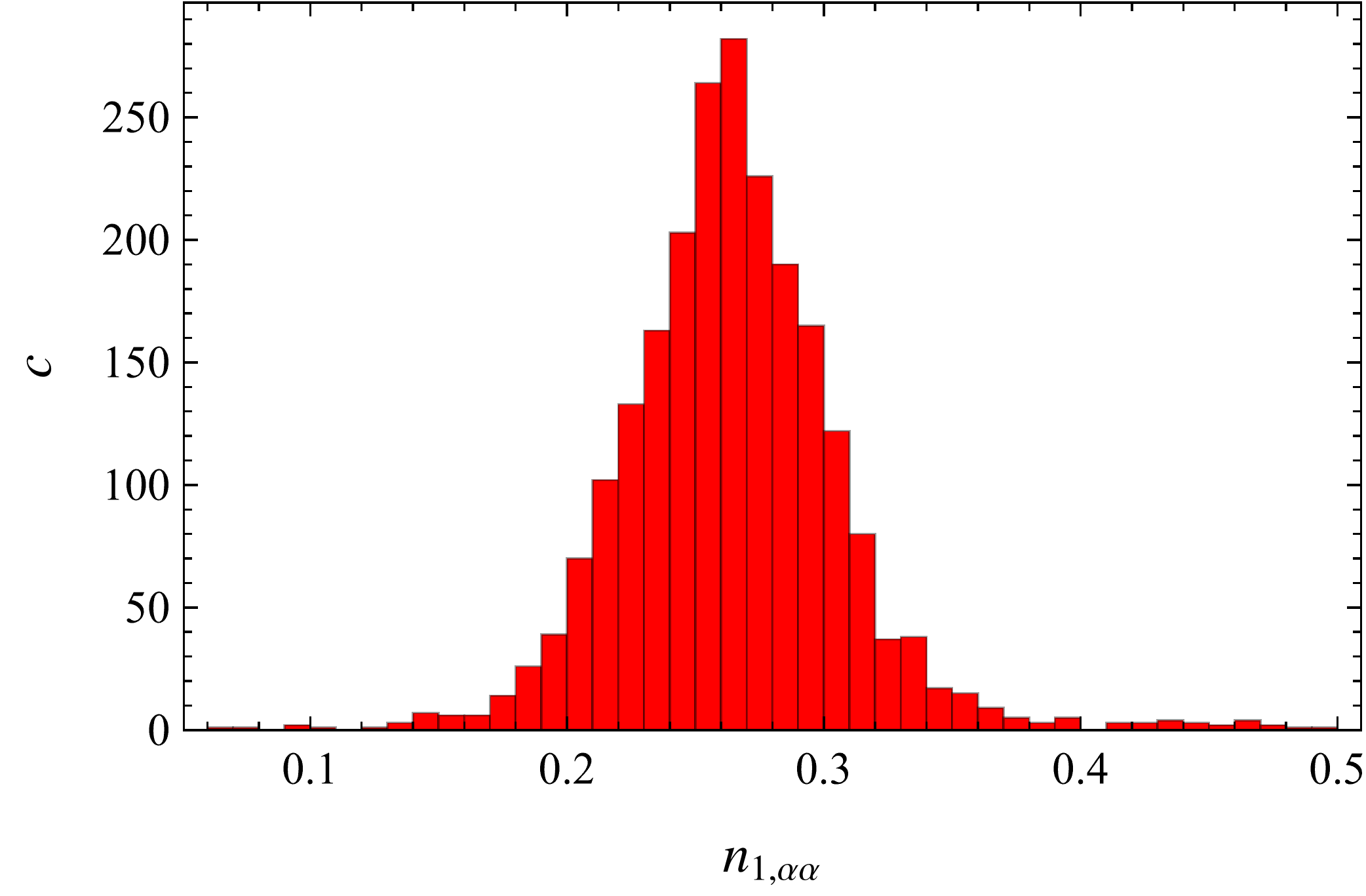}
        \label{fig_hist_IN_N_08}
    \end{subfigure}
    \caption{\small Thermalization despite correlation. The subfigures display: (\subref{fig_DLT_n_0_aa_IO_N_08}) a plot of $\Delta_{n_{1,\alpha \alpha}}$ over $E_{\alpha}$ for $N=8$, (\subref{fig_DLT_C_a_sq_IO_N_08}) a plot of $\Delta_{C_{\alpha}^2}$ over $E_{\alpha}$ for $N=8$, (\subref{fig_DLT_n_0_DLT_C_IO_N_08}) a plot of $\Delta_{C_{\alpha}^2}$ over $\Delta_{n_{1,\alpha \alpha}}$ for $N=8$, (\subref{fig_hist_N_08}) a histogram of count $c$ over $n_{1,\alpha \alpha}$ and $C_{\alpha}^2$ for $N=8$, (\subref{fig_hist_ALL_N_08}) a histogram of count $c$ over $n_{1,\alpha \alpha}$ for $N=8$, (\subref{fig_hist_IN_N_08}) a histogram of count $c$ over $n_{1,\alpha \alpha}$ within the microcanonical energy window for $N=8$. In subfigures (\subref{fig_DLT_n_0_aa_IO_N_08}), (\subref{fig_DLT_C_a_sq_IO_N_08}) and (\subref{fig_DLT_n_0_DLT_C_IO_N_08}) the points within the microcanonical energy shell $\overline{E} \pm \sigma_{E,\text{q}}$ are shown in red, all other points are shown in blue. In subfigure (\subref{fig_hist_N_08}) the bins for points within the microcanonical energy shell are shown in red and the bins for all other points are shown in blue. The bins for identical values of $n_{1,\alpha \alpha}$ and $C_{\alpha}^2$ are stacked. Likewise, in subfigure (\subref{fig_hist_ALL_N_08}) the bins for points within the microcanonical energy shell are shown in red and the bins for all other points are shown in blue. Here the bins over identical $n_{1,\alpha \alpha}$ are stacked as well.}
\end{figure}

Figures~\ref{fig_DLT_n_0_aa_IO_N_08}, \ref{fig_DLT_C_a_sq_IO_N_08} and~\ref{fig_DLT_n_0_DLT_C_IO_N_08} display exemplary plots for $N=8$ of $\Delta_{n_{1,\alpha\alpha}}$ over $E_{\alpha}$, $\Delta_{C_{\alpha}^2}$ over $E_{\alpha}$ and $\Delta_{C_{\alpha}^2}$ over $\Delta_{n_{1,\alpha\alpha}}$, respectively. The corresponding histograms of count $c$ over $n_{1,\alpha\alpha}$ and $C_{\alpha}^2$, count $c$ over $n_{1,\alpha\alpha}$, and count $c$ over $n_{1,\alpha\alpha}$ for points within the microcanonical energy window are shown in Figs.~\ref{fig_hist_N_08}, \ref{fig_hist_ALL_N_08} and~\ref{fig_hist_IN_N_08}, respectively. These results suggest that the two sets of fluctuations, $\Delta_{C_{\alpha}^2}$ and $\Delta_{n_{1,\alpha\alpha}}$, are correlated. Specifically, we observe that larger values of $C_{\alpha}^2$ correspond to larger values of $n_{1,\alpha\alpha}$. We find an analogous dependence between $C_{\alpha}^2$ and $n_{1,\alpha\alpha}$ for other values of $N$.

We also comment on the tails of the distribution of the diagonal matrix elements $n_{1,\alpha\alpha}$ in Figs.~\ref{fig_hist_N_08} and~\ref{fig_hist_ALL_N_08}. These diminish at different rates as $n_{1,\alpha\alpha}$ diverges from $n_{1,\text{av}}$. Specifically, we remark that as $n_{1,\alpha\alpha}$ moves away from $n_{1,\text{av}}$, the left tail for $n_{1,\alpha\alpha} < n_{1,\text{av}}$ decreases slower than the right tail for $n_{1,\alpha\alpha} > n_{1,\text{av}}$. At the same time, the distribution of the $n_{1,\alpha\alpha}$-s within the microcanonical energy window (see Fig.~\ref{fig_hist_IN_N_08}) appears to follow a more symmetrical distribution. These characteristics are crucial for the thermalization mechanism that we introduce in section~\ref{subsec_new_therm_mech}.

We now analyze the two sets of fluctuations more closely. Specifically, we study numerically how the correlation between them depends on $N$. Note that a necessary requirement of both the mechanism (i) of~\cite{Rigol_2008} and the ERH is that the fluctuations of the $n_{1,\alpha \alpha}$-s and the $C_{\alpha}^2$-s are \emph{absolutely} uncorrelated. Namely, any amount of correlation in the two sets of fluctuations would render these two mechanisms inapplicable, provided the correlation does not vanish in the TDL.

\begin{figure}[htbp]
    \centering
    \begin{subfigure}[b]{0.32\textwidth}
        \caption{\footnotesize Blomqvist $\beta$}
        \includegraphics[width=\textwidth]{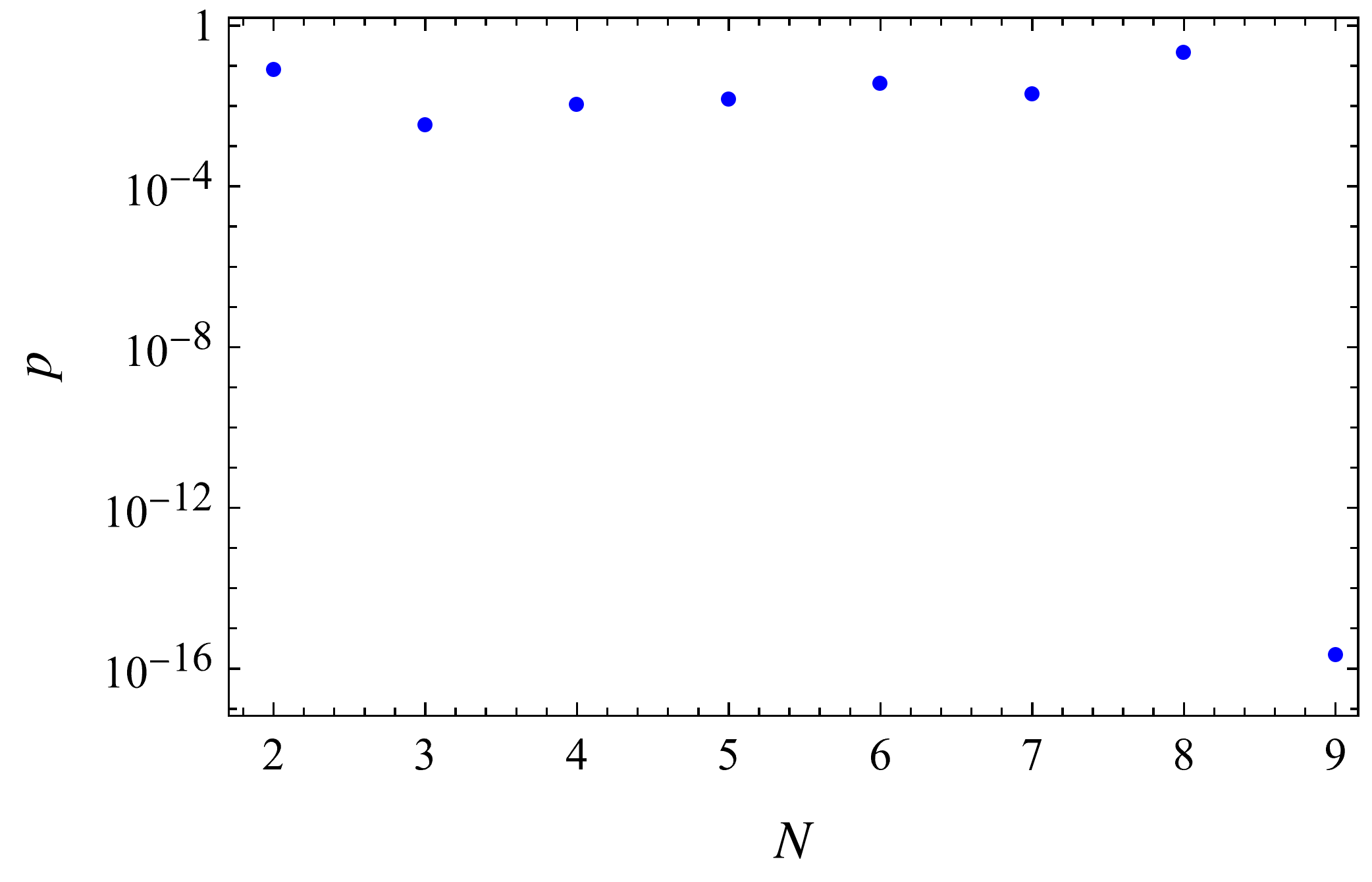}
        \centering
        \label{fig_BL}
    \end{subfigure}
    \begin{subfigure}[b]{0.32\textwidth}
    	\caption{\footnotesize Goodman-Kruskal $\gamma$}
        \includegraphics[width=\textwidth]{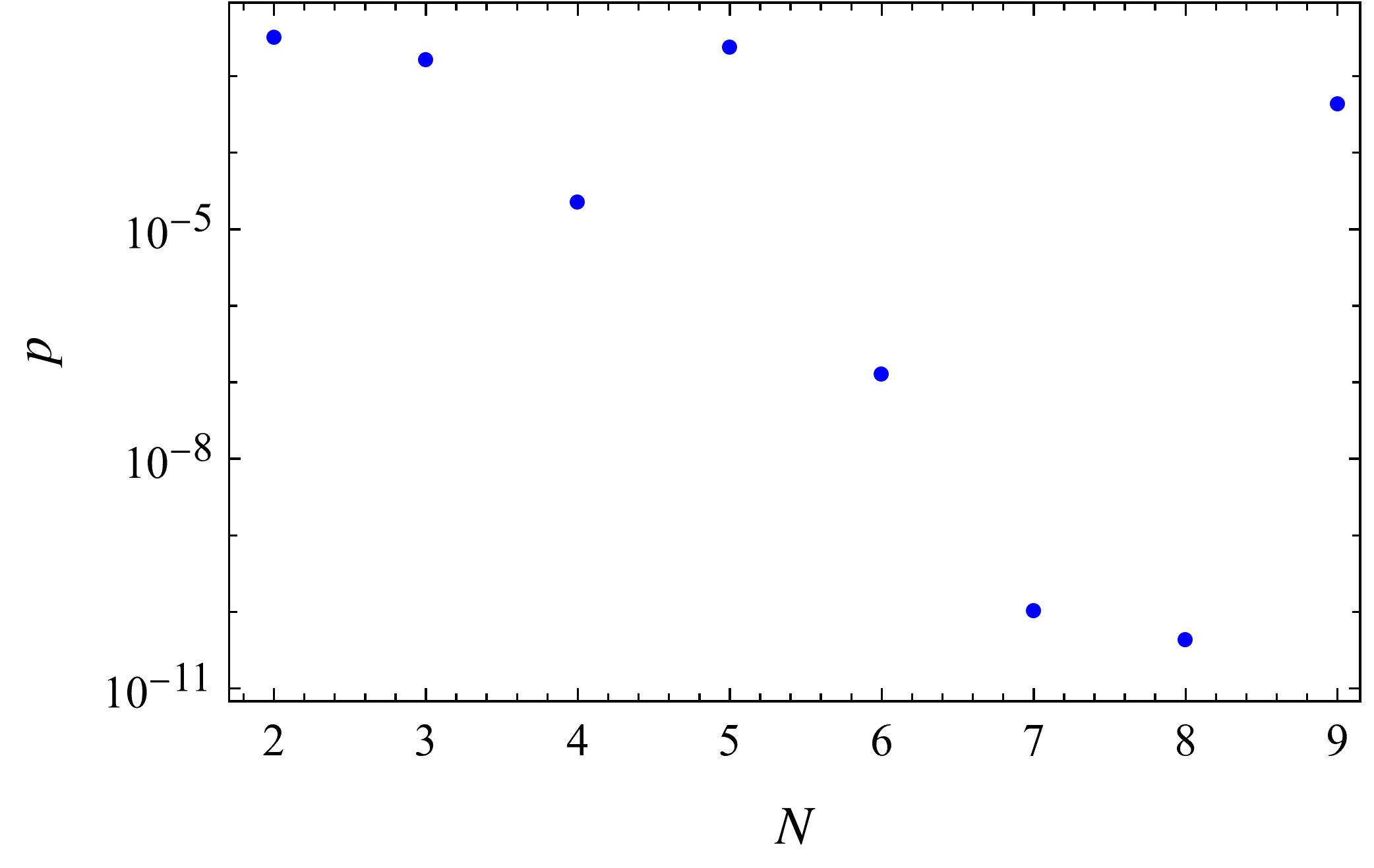}
        \label{fig_GO}
    \end{subfigure}
    \begin{subfigure}[b]{0.32\textwidth}
    	\caption{\footnotesize Hoeffding $\mathcal{D}$}
        \includegraphics[width=\textwidth]{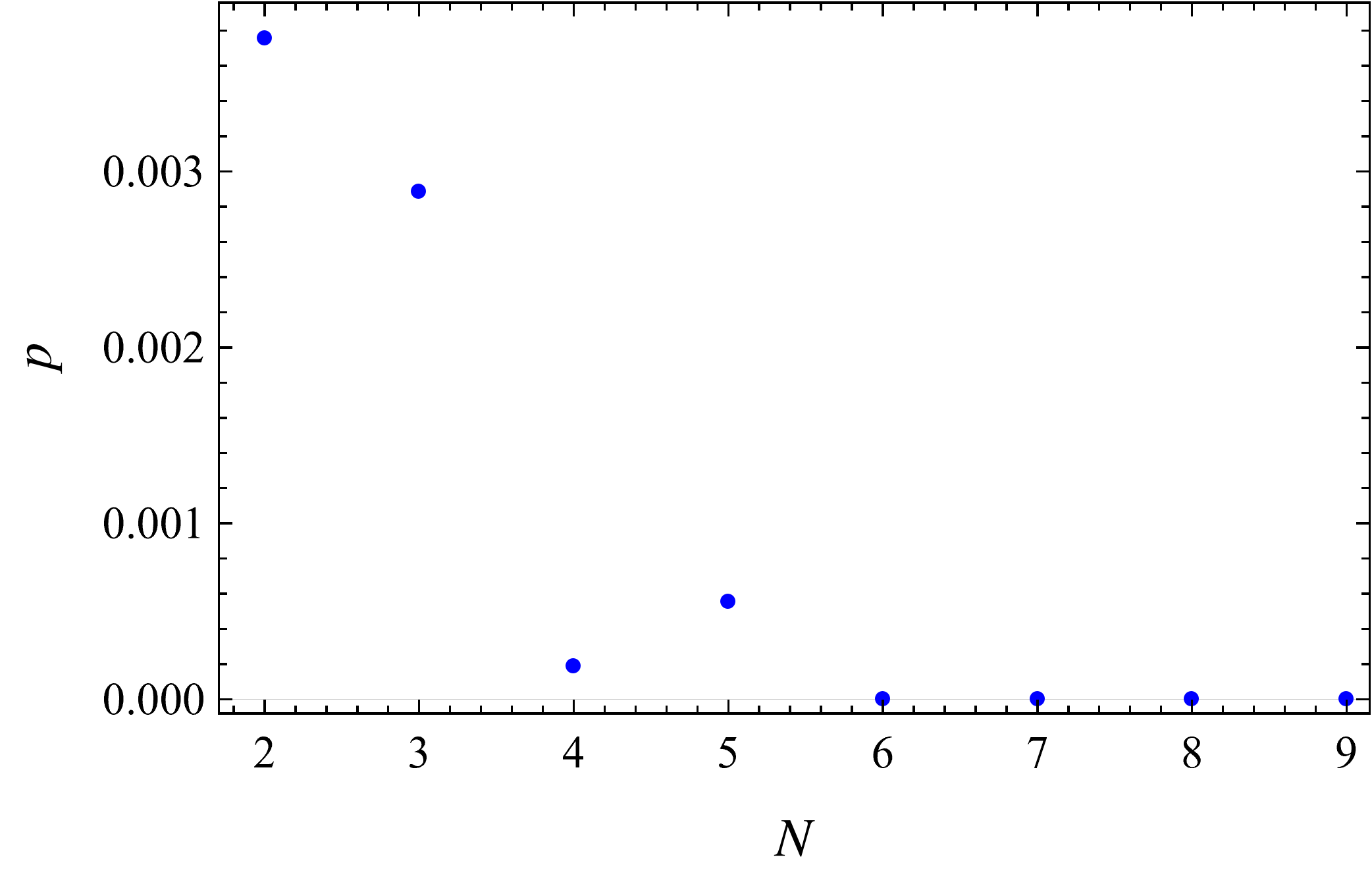}
        \label{fig_HO}
    \end{subfigure}
    \begin{subfigure}[b]{0.32\textwidth}
    	\caption{\footnotesize Kendall $\tau$}
        \includegraphics[width=\textwidth]{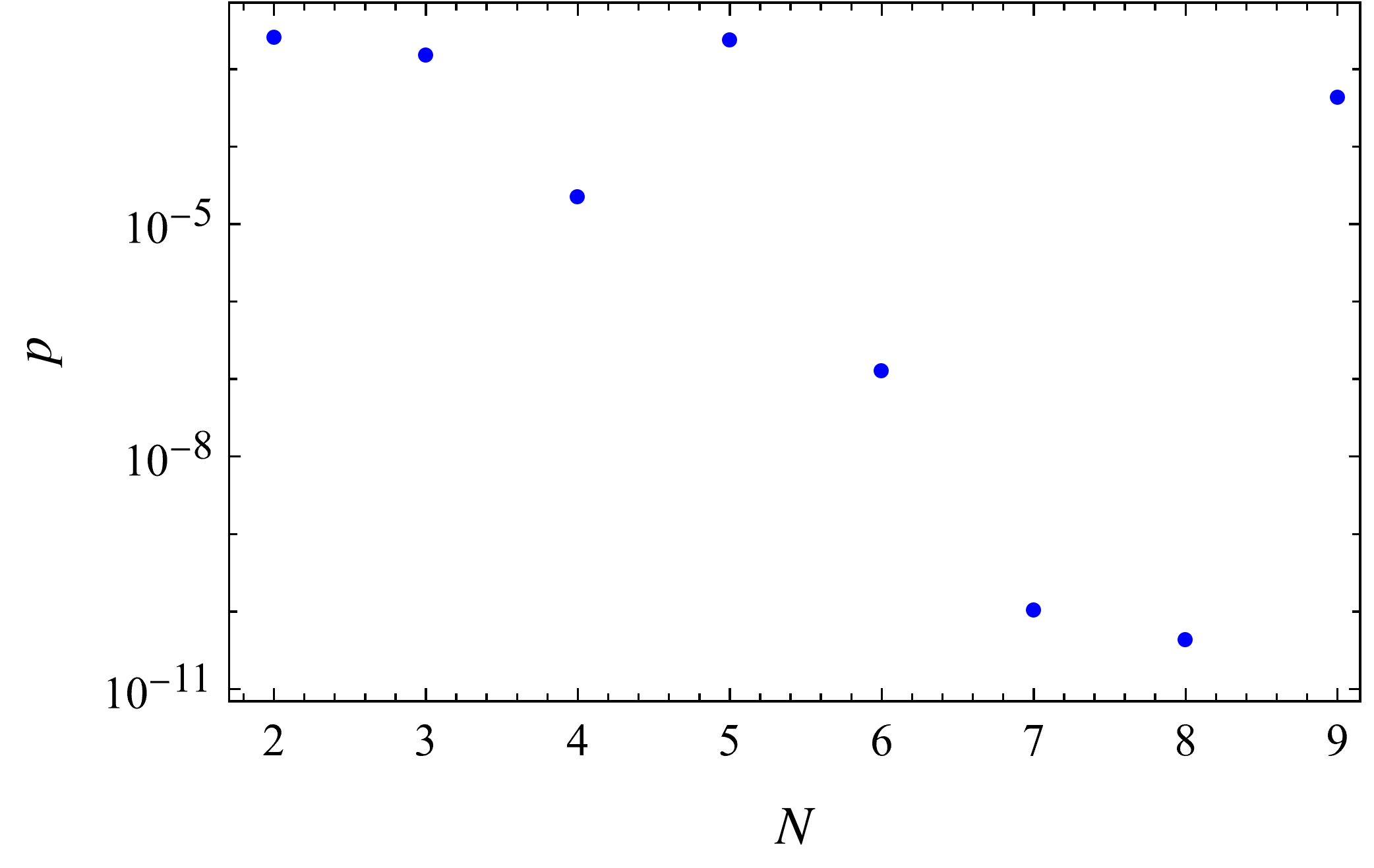}
        \label{fig_KE}
    \end{subfigure}
    \begin{subfigure}[b]{0.32\textwidth}
    	\caption{\footnotesize Spearman Rank}
        \includegraphics[width=\textwidth]{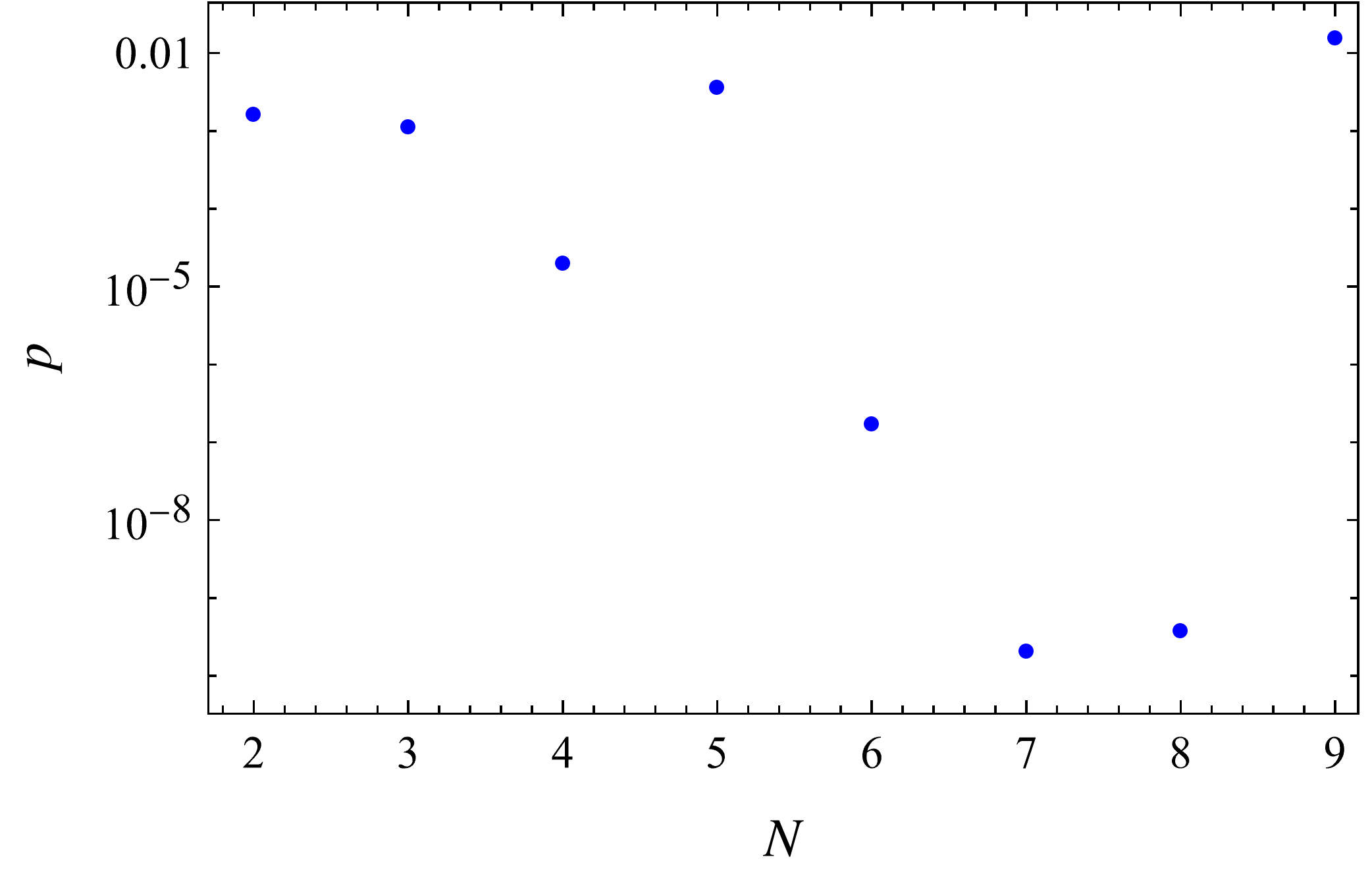}
        \label{fig_SP}
    \end{subfigure}
    \caption{\small Independence tests for $\Delta_{n_{1,\alpha \alpha}}$ and $\Delta_{C_{\alpha}^2}$. The subfigures show $p$ over $N$ for the following independence tests: (\subref{fig_BL}) Blomqvist $\beta$, (\subref{fig_GO}) Goodman-Kruskal $\gamma$, (\subref{fig_HO}) Hoeffding $\mathcal{D}$, (\subref{fig_KE}) Kendall $\tau$, (\subref{fig_SP}) Spearman Rank. The Pearson Correlation, Pillai Trace and Wilks $\mathcal{W}$ independence tests were not applicable to the data. The $p$-scale in subfigures (\subref{fig_BL}), (\subref{fig_GO}), (\subref{fig_KE}) and (\subref{fig_SP}) is logarithmic. The results of $p$ for the independence test in (\subref{fig_HO}) for $N \geq 6$ were below the numerical accuracy and were returned as zero. Except for $p \approx 0.08$ at $N=2$ and $p \approx 0.21$ at $N=8$ in (\subref{fig_BL}), all probabilities $p$ are below the significance level $\alpha=0.05$.}
    \label{fig_hyp_tst}
\end{figure}

To examine this, we conduct a hypothesis test on the two sets of fluctuations, $\Delta_{n_{1,\alpha \alpha}}$ and $\Delta_{C_{\alpha}^2}$, for each value of $N$. Here, the null hypothesis $H_0$ is that these data sets are independent, and the alternative hypothesis $H_{\text{a}}$ is that they are mutually dependent. We set the significance level $\alpha$ equal to $0.05$. The output of each independence test is a probability value $p$. If $p < \alpha$, the null hypothesis is rejected at that significance level $\alpha$. Figure~\ref{fig_hyp_tst} shows our findings.

We find that except for the results of the Blomqvist $\beta$ independence test at $N=2$ and $N=8$ (see Fig.~\ref{fig_BL}), all probabilities $p$ are below the significance level $\alpha=0.05$. Therefore, at the significance level of $\alpha=0.05$, the results of the independence tests indicate that the null hypothesis that the two sets of fluctuations $\Delta_{n_{1,\alpha \alpha}}$ and $\Delta_{C_{\alpha}^2}$ are uncorrelated is to be rejected.

Furthermore, in the independence tests in Figs.~\ref{fig_GO}, \ref{fig_HO}, \ref{fig_KE} and~\ref{fig_SP} the $p$-values for even $N$ (i.e.\ where $N_m$ does not need to be adjusted by rounding down to the nearest integer) exhibit the trend of decreasing $p$ for increasing $N$. This suggests that for larger $N$ the two sets of fluctuations remain correlated. Specifically, together with the findings in Fig.~\ref{fig_DLT_n_0_DLT_C_IO_N_08}, this indicates that the bias of the $C_{\alpha}^2$-s towards the $n_{1,\alpha \alpha}$-s with a greater deviation from the mean persists in the TDL. In particular, this suggests that the conditions of the weak ETH are not fulfilled.

Our results indicate that the fluctuations of the $n_{1,\alpha \alpha}$-s and those of the $C_{\alpha}^2$-s remain significantly correlated in the TDL. This suggests that thermalization occurs neither via the mechanism (i) of~\cite{Rigol_2008}, nor via the ERH, but instead via a different mechanism. We propose the corresponding mechanism in section~\ref{subsec_new_therm_mech}.

\section{Discussion} \label{sec_disc}

\subsection{A new thermalization mechanism} \label{subsec_new_therm_mech}

We recall from section~\ref{subsec_therm}, that there are only two conditions necessary for the thermalization of a few-body observable $\hat{A}$ in an isolated many-body system evolved from a non-equilibrium initial state. These are
\begin{itemize}
\item[(i)]
\begin{equation}
\sum\limits_{\alpha} |C_{\alpha}|^2 A_{\alpha\alpha} = \dfrac{1}{\mathcal{N}_{\sigma_{\text{w}}}} \sum\limits_{\substack{\alpha \\ |\overline{E} - E_{\alpha}| < \sigma_{\text{w}}}} A_{\alpha \alpha} ~ \text{for an appropriate energy window} ~ 2\sigma_{\text{w}} \text{, and}
\end{equation}
\item[(ii)]
\begin{equation}
\sigma_t = \left[ \sum\limits_{\substack{\alpha, \beta \\ \alpha \neq \beta}} |C_{\alpha}|^2 |C_{\beta}|^2 |A_{\alpha\beta}|^2 \right]^{1/2} ~ \text{is small at most later times,}
\end{equation}
\end{itemize}
where $\mathcal{N}_{\sigma_{\text{w}}}$ is the number of eigenstates $\ket{\alpha}$ whose eigenenergies $E_{\alpha}$ are within the energy window $(\overline{E}-\sigma_{\text{w}},\overline{E}+\sigma_{\text{w}})$. The particular details of how these conditions are fulfilled are not relevant to the ultimate consequence, which is thermalization. The diagonal matrix elements $A_{\alpha\alpha}$ and/or the squared magnitudes of the coefficients $|C_{\alpha}|^2$ may exhibit fluctuations. Irrespective of how large or correlated these fluctuations are, the observable may still thermalize, provided the conditions are met.

The ETH is an elegant thermalization mechanism with simple conditions on the coefficients and the matrix elements. Although the ETH conditions are sufficient for a few-body observable to thermalize, they do not represent a necessary set of conditions. Multiple alternative thermalization mechanisms exist in the literature, such as the ones mentioned in section~\ref{sec_test_other_mech} and many others. The findings of this work suggest a new variation of the conditions for thermalization. Specifically, there may exist correlations between the quantities from which the infinite-time average of the expectation value of the observable is computed. Despite this, the infinite-time average of the observable can be equal to its statistical ensemble average.

Let us recapitulate the progress so far. In section~\ref{subsec_therm_test_num_an} we studied the time-evolution properties of the few-body observable $\hat{n}_1$ within the model (\ref{eqn_model}) evolved from the non-equilibrium initial state (\ref{eqn_in_st}). We found indications that $\hat{n}_1$ thermalizes in the TDL according to the definitions (i) and (ii) in section~\ref{subsubsec_therm_cond}. The results in section~\ref{sec_ETH_test} suggest that the ETH condition (0) on the initial state and the ETH condition (2) on the magnitudes of the off-diagonal matrix elements $|n_{1,\alpha\beta ; \alpha \neq \beta}|$ (see section~\ref{subsubsec_ETH_cond} for definitions) are both satisfied. The latter finding is consistent with the observed indication that the thermalization condition (ii) from section~\ref{subsubsec_therm_cond} is fulfilled.

However, the results of section~\ref{sec_ETH_test} also suggest that the ETH condition (1) from section~\ref{subsubsec_ETH_cond} is not satisfied. Specifically, we find indications that the diagonal matrix elements $n_{1,\alpha\alpha}$ do not vary smoothly with $E_{\alpha}$, and that the maximal difference between neighboring diagonal elements within the microcanonical energy window approaches a value of $\gtrsim 1.04 n_{1,\text{mc}}$ in the TDL. Furthermore, we find indications that the multiple thermalization mechanisms listed in sections~\ref{sec_test_other_mech} and~\ref{sec_TDC} are also not applicable.

In section~\ref{sec_TDC} we find indications that the fluctuations of $n_{1,\alpha \alpha}$ and $C_{\alpha}^2$ remain correlated in the TDL. Furthermore, the results of section~\ref{subsec_therm_test_num_an} indicate that $\overline{n}_1 = n_{1,\text{mc}}$ in the TDL. These two observations are reconciled as follows: For $\overline{n}_1$ on the left-hand side, the fluctuations are large but are correlated in such a way that a few $n_{1,\alpha\alpha}$-s with $n_{1,\alpha\alpha} > n_{1,\text{av}}$ are biased towards by large $C_{\alpha}^2$-s, while many $n_{1,\alpha\alpha}$-s with $n_{1,\alpha\alpha} < n_{1,\text{av}}$ have smaller weights $C_{\alpha}^2$ (see also Figs.~\ref{fig_hist_N_08} and~\ref{fig_hist_ALL_N_08}). That is, within the weighted sum for $\overline{n}_1$, the many small $n_{1,\alpha\alpha}$-s with small weights collectively balance out the few large $n_{1,\alpha\alpha}$-s with large weights. On the right-hand side, the unweighted average of the $n_{1,\alpha\alpha}$-s within the microcanonical energy window is $n_{1,\text{mc}}$ (see also Fig.~\ref{fig_hist_IN_N_08}). This satisfies the thermalization condition (i) from section~\ref{subsubsec_therm_cond}.

We emphasize that our findings indicate that the following features persist in the TDL: First, large fluctuations between the neighboring $n_{1,\alpha\alpha}$-s. Second, a few $n_{1,\alpha\alpha}$-s with $n_{1,\alpha\alpha} > n_{1,\text{av}}$ biased towards by large $C_{\alpha}^2$-s. Third, many $n_{1,\alpha\alpha}$-s with $n_{1,\alpha\alpha} < n_{1,\text{av}}$ corresponding to small $C_{\alpha}^2$-s. That is, in the large-$N$ limit the distribution of the $n_{1,\alpha\alpha}$-s does not become symmetric with a correspondingly unbiased distribution for the $C_{\alpha}^2$-s. Therefore, $\overline{n}_1 = n_{1,\text{mc}}$ is satisfied precisely because of the particular distributions and correlations of the $n_{1,\alpha\alpha}$-s and the $C_{\alpha}^2$-s.

To reiterate: the thermalization condition (i) requires only that $\overline{n}_1 = n_{1,\text{mc}}$. For this equality to hold, the specific distributions of the $n_{1,\alpha\alpha}$-s and the $C_{\alpha}^2$-s are not required to fulfill any additional constraints. In particular, the distributions of the two quantities can exhibit large fluctuations and be arbitrarily correlated, as long as $\overline{n}_1 = n_{1,\text{mc}}$.

Variations of the presented mechanism are also possible: First, instead of the above, a few $n_{1,\alpha\alpha}$-s with $n_{1,\alpha\alpha} < n_{1,\text{av}}$ biased towards by large $C_{\alpha}^2$-s, and many $n_{1,\alpha\alpha}$-s with $n_{1,\alpha\alpha} > n_{1,\text{av}}$ and small $C_{\alpha}^2$-s. Second, a symmetric version, with the $n_{1,\alpha\alpha}$-s that deviate strongly from $n_{1,\text{av}}$ corresponding to large $C_{\alpha}^2$-s. This version could allow for large fluctuations of a substantial fraction of the $n_{1,\alpha\alpha}$-s. Moreover, the fraction of the non-thermal states could be allowed to be non-zero in the TDL. Nevertheless, $\overline{n}_1 = n_{1,\text{mc}}$ could hold with suitably correlated distribution of the $C_{\alpha}^2$-s. In this case, not only $\delta_{1,\text{max;mc}}$, but also $\delta_1$ and $\delta_{1,\text{mc}}$ would be non-zero in the TDL.

An interesting question is whether a few-body observable can thermalize via the proposed mechanism within a finite-size isolated many-body quantum system of $D$ degrees of freedom. That is, can the fluctuations of the diagonal elements and the coefficients-squared be arranged in such a way that a few-body observable thermalizes for finite values of $D$? This also raises the question of how much control we have over the correlations of the fluctuations. For example, in the black hole scaling regime of the model (\ref{eqn_model}), there are $N(2N-1) \sim \mathcal{O}(N^2)$ couplings that we control. However, the Hilbert space dimension grows exponentially with $N$. Achieving complete control of the fluctuations in the large-$N$ limit is therefore highly improbable.

As a supplementary remark, we point out a recent work~\cite{Foini_2024}, which studied how the correlations between the initial state of the system and the off-diagonal elements of the observable determine the non-equilibrium dynamics.

\subsection{Applications}

Our findings are relevant across several domains. Here we outline the implications of our results.

The system of enhanced memory capacity studied in this work was formulated to model the information-processing characteristics of a black hole. The results of the present work indicate that the occupation numbers of the memory-storing modes in this prototype system thermalize. Moreover, our findings indicate a microscopic mechanism through which this occurs. We do not claim that the information within black holes undergoes inner entanglement via the proposed mechanism: Our results apply to the particular model considered in this work. Nevertheless, this system exhibits interesting time-evolution features and represents a valuable prospect for future studies.

Furthermore, there is an additional practical aspect to the model (\ref{eqn_model}). As previously suggested in \cite{Dvali_2019_2} among other works, it may be feasible to study such a system in ultracold atom experiments under laboratory conditions~\cite{Bloch_2005, Bloch_2008, Bloch_2012, Wienand_2023}. The experimental realization of the model may allow for tests of its predicted information-processing characteristics. Furthermore, an actualization of the model with ultracold atoms would enable the implementation of its intended regime of operation; $N \gg 1$. In addition, the model (\ref{eqn_model}) is similar to the Bose-Hubbard model (BHM)~\cite{Gersch_1962, Fisher_1989}, which has already been successfully realized in many experiments.

\section{Conclusion} \label{sec_concl}

In the present paper we have studied the long-time behavior of certain few-body observables within the isolated quantum many-body system (\ref{eqn_model})~\cite{Dvali_2020} evolved from the non-equilibrium initial state (\ref{eqn_in_st}). This system exhibits an enhanced capacity to store information and has been designed to model the information-processing characteristics of a black hole. In a black hole-like scaling regime of the model parameters, we have found indications that the chosen few-body observables within the system thermalize in the TDL according to the definitions (i) and (ii) in section~\ref{subsubsec_therm_cond}. Specifically, our results suggest that the infinite-time averages $\overline{n}_i = \overline{\braket{\hat{n}_i(t)}}$ of the time-evolved expectation values $\braket{\hat{n}_i(t)}$ of the occupation number operators $\hat{n}_i$ of the information-carrying modes approach their corresponding microcanonical ensemble averages $n_{i,\text{mc}}$ in the large system-size limit $N \to \infty$. Our findings are based on the large-$N$ extrapolation of numerically obtained results for finite-size realizations of the system.

Furthermore, our observations suggest that this thermalization occurs via a novel mechanism. The fundamental new aspect is that the fluctuations of both the diagonal matrix elements $A_{\alpha\alpha}$ of a few-body observable $\hat{A}$ in the eigenstate basis of the model $\{\ket{\alpha}\}$ and the squared magnitudes of the coefficients $|C_{\alpha}|^2 = |\braket{\alpha | \text{in}} |^2$ remain not only significant, but also mutually dependent in the TDL.

A necessary condition for the thermalization of the observable $\hat{A}$ (see condition (i) in section~\ref{subsubsec_therm_cond}) is the equality of the corresponding infinite-time and microcanonical ensemble averages, $\overline{A}=A_{\text{mc}}$. Expressed in the diagonal ensemble, the infinite time-average $\overline{A}=\overline{\braket{\hat{A}(t)}}=\sum\limits_{\alpha}|C_{\alpha}|^2 A_{\alpha\alpha}$ is determined by the $A_{\alpha\alpha}$-s and the $|C_{\alpha}|^2$-s. The novelty of the proposed mechanism is the following: While on the right-hand side of $\overline{A}=A_{\text{mc}}$ the unweighted average of the $A_{\alpha\alpha}$-s within the microcanonical energy window is $A_{\text{mc}}$, on the left-hand side the $A_{\alpha\alpha}$-s and the $|C_{\alpha}|^2$-s fluctuate considerably, but are also correlated. Therefore, the $A_{\alpha\alpha}$-s are paired with non-random weights $|C_{\alpha}|^2$-s. However, for the condition $\overline{A}=A_{\text{mc}}$ the amount of bias within the sum for $\overline{A}$ is irrelevant.

Specifically, in our numerical simulations we observe that within the sum for the infinite-time average the many smaller-than-average diagonal elements with small weights collectively counter-balance the few larger-than-average diagonal elements with large weights. The independence tests performed on the two sets of fluctuations indicate that these are correlated: At a significance level of $0.05$, the null hypothesis that these quantities are independent is rejected in favor of the alternative hypothesis that they are mutually dependent.

A realization of this system in ultracold atom experiments may allow for further studies of its information-processing features.


\section*{Acknowledgments}

This work has received support from the French State managed by the National Research Agency under the France 2030 program with reference ANR-22-PNCQ-0002. We would like to thank Mari-Carmen Ba\~{n}uls and Maxim Olshanii for valuable discussions.


\begin{appendix}

\section*{Appendix} 

\begin{table}[H]
\centering
\begin{adjustbox}{width=\textwidth}
\begin{tabular}{l|l|l|l|l|l|l|l|l}
Quantity & $\overline{R}^2$ & $\text{RMSE}$ & $a$ & $\sigma_{a}$ & $b$ & $\sigma_{b}$ & $c$ & $\sigma_{c}$ \\
\hline
$^* ~ \overline{n}_1(N)=a+b \exp(cN)$ & $9.966 \times 10^{-1}$ & $1.96 \times 10^{-2}$ & $2.7 \times 10^{-1}$ & $1.5 \times 10^{-1}$ & $1.5 \times 10^{-1}$ & $1.1 \times 10^{-1}$ & $-1.4 \times 10^{-1}$ & $3.2 \times 10^{-1}$ \\
$\overline{n}_1(N)=a+b N^{-1/2}$ & $9.968 \times 10^{-1}$ & $1.91 \times 10^{-2}$ & $2.58 \times 10^{-1}$ & $2.8 \times 10^{-2}$ & $1.78 \times 10^{-1}$ & $6.1 \times 10^{-2}$ & $\text{n/a}$ & $\text{n/a}$ \\
$^* ~ n_{1,\text{mc}}(N)=a+b \exp(cN)$ & $9.998 \times 10^{-1}$ & $3.47 \times 10^{-3}$ & $2.53 \times 10^{-1}$ & $1.5 \times 10^{-2}$ & $7.9 \times 10^{-2}$ & $1.0 \times 10^{-2}$ & $-2.5 \times 10^{-1}$ & $1.6 \times 10^{-1}$ \\
$n_{1,\text{mc}}(N)=a+b N^{-1}$ & $9.997 \times 10^{-1}$ & $4.56 \times 10^{-3}$ & $2.538 \times 10^{-1}$ & $4.7 \times 10^{-3}$ & $9.5 \times 10^{-2}$ & $1.6 \times 10^{-2}$ & $\text{n/a}$ & $\text{n/a}$ \\
$n_{1,\text{mc}}(N)=a+b N^{-1/2}$ & $9.999 \times 10^{-1}$ & $3.33 \times 10^{-3}$ & $2.273 \times 10^{-1}$ & $6.3 \times 10^{-3}$ & $1.04 \times 10^{-1}$ & $1.2 \times 10^{-2}$ & $\text{n/a}$ & $\text{n/a}$ \\
$\sigma_{1,t}(N)=a+b \exp(cN)$ & $9.96 \times 10^{-1}$ & $5.45 \times 10^{-3}$ & $8.9 \times 10^{-3}$ & $9.9 \times 10^{-3}$ & $4.10 \times 10^{-1}$ & $4.7 \times 10^{-2}$ & $-4.70 \times 10^{-1}$ & $7.4 \times 10^{-2}$ \\
$\sigma_{E,\text{q}} / N (N)=a+b N^c$ & $9.999 \times 10^{-1}$ & $3.05 \times 10^{-3}$ & $-2.3 \times 10^{-2}$ & $1.3 \times 10^{-2}$ & $1.250$ & $1.1 \times 10^{-2}$ & $-9.00 \times 10^{-1}$ & $3.3 \times 10^{-2}$ \\
$^* ~ \mathcal{N}_{\sigma_{E,\text{q}}}(N)=a+b \exp(cN)$ & $9.75 \times 10^{-1}$ & $248$ & $-107$ & $160$ & $6.7$ & $7.4$ & $7.1 \times 10^{-1}$ & $1.2 \times 10^{-1}$ \\
$\mathcal{N}_{\sigma_{E,\text{q}}}(N)=a+b N^c$ & $9.80 \times 10^{-1}$ & $221$ & $-34$ & $118$ & $9 \times 10^{-3}$ & $1.6 \times 10^{-2}$ & $5.93$ & $8.5 \times 10^{-1}$ \\
$n_{1,\text{av}}(N)=a+b N^{-1}$ & $1.$ & $4.45 \times 10^{-16}$ & $2.5 \times 10^{-1}$ & $5.6 \times 10^{-16}$ & $-2.5 \times 10^{-1}$ & $2.6 \times 10^{-15}$ & $\text{n/a}$ & $\text{n/a}$ \\
$\delta_1(N)=a+b \exp(cN)$ & $9.98 \times 10^{-1}$ & $1.74 \times 10^{-2}$ & $-2 \times 10^{-2}$ & $7.9 \times 10^{-1}$ & $1.7$ & $1.9$ & $-2.9 \times 10^{-1}$ & $6.6 \times 10^{-1}$ \\
$\delta_{1,\text{mc}}(N)=a+b \exp(cN)$ & $9.99 \times 10^{-1}$ & $1.31 \times 10^{-2}$ & $-5 \times 10^{-1}$ & $8.2$ & $1.5$ & $6.9$ & $-1.0 \times 10^{-1}$ & $8.9 \times 10^{-1}$ \\
$^* ~ \delta_{\substack{1,\text{max} \\ \text{mc}}}(N)=a+b N^c$ & $9.99995 \times 10^{-1}$ & $3.00 \times 10^{-3}$ & $1.0442$ & $9.2 \times 10^{-3}$ & $1.620$ & $4.1 \times 10^{-2}$ & $-1.451$ & $5.5 \times 10^{-2}$ \\
$\delta_{\substack{1,\text{max} \\ \text{mc}}}(N)=a+b\exp(cN)$ & $9.9998 \times 10^{-1}$ & $5.44 \times 10^{-3}$ & $1.1157$ & $6.3 \times 10^{-3}$ & $1.831$ & $9.3 \times 10^{-2}$ & $-6.29 \times 10^{-1}$ & $2.8 \times 10^{-2}$ \\
$|n_{1,\alpha \beta ; \alpha \neq \beta}|_{\text{av}}(N)=a+b \exp(cN)$ & $9.94 \times 10^{-1}$ & $2.49 \times 10^{-3}$ & $-5 \times 10^{-4}$ & $2.8 \times 10^{-3}$ & $2.66 \times 10^{-1}$ & $4.0 \times 10^{-2}$ & $-6.54 \times 10^{-1}$ & $8.2 \times 10^{-2}$
\end{tabular}
\end{adjustbox}
\caption{\small Best obtained fit functions. For each fit, both the coefficients of determination $\overline{R}^2$ and the unbiased root-mean-square errors (RMSE) are adjusted for the respective number of free fit-model parameters. The standard error of a fit parameter $p_{\text{fit}} \in \{ a, b, c \}$ is denoted by $\sigma_{p_{\text{fit}}}$. If multiple fits are performed for a quantity, the fits displayed in the figures throughout the paper are marked with a ``$^*$''. For $\overline{n}_1(N)$, the outlier at $N=3$ is excluded from the corresponding fits. The fits for $n_{1,\text{mc}}(N)$ are performed only over the points with even $N$. The fit for $n_{1,\text{av}}(N)$ is performed only over the points with odd $N$. The fit for $\delta_1(N)$ is a weighted fit with weights $\sigma_1^{-2}(N)$ and is performed only over the points with even $N$. The fit for $\delta_{1,\text{mc}}(N)$ is a weighted fit with weights $\sigma_{1,\text{mc}}^{-2}(N)$ and is performed only over the points with even $N$. The fits for $\delta_{1,\text{max}; \text{mc}}(N)$ are performed only over the points with even $N$.}
\label{tab_fit}
\end{table}

\end{appendix}


%
%
%
\end{document}